\documentclass[12pt,preprint]{aastex}
\begin{document}
\title{Blue Stragglers in Galactic Open Clusters and the Integrated 
Spectral Energy Distributions}

\author{Y. Xin and L. Deng}
\affil{National Astronomical Observatories, Chinese Academy of Sciences,
Beijing 100012}
\email{xinyu@bao.ac.cn, licai@bao.ac.cn}

\begin{abstract}
Synthetic integrated spectral properties of the old Galactic open clusters
are studies in this work, where twenty-seven Galactic open clusters of 
ages $\geq$ 1Gyr are selected as the working sample. 
Based on the photometric observations of these open clusters,
synthetic integrated spectrum has been made for the stellar population 
of each cluster. The effects of blue straggler stars (BSSs) on
the conventional simple stellar population (SSP) model are analyzed on an
individual cluster base. It is shown that the BSSs, whose holding positions 
in the color-magnitude diagrams (CMDs) cannot be predicted by the current 
single-star evolution theory, present significant modifications to the 
integrated properties of theoretical SSP model. 
The synthesized integrated spectral energy distributions (ISEDs) of our sample 
clusters are dramatically different from the SSPs based on isochrone 
only. The BSSs corrected ISEDs of stellar populations show systematic
enhancements towards shorter wavelength in the spectra. 
When measured with wide-band colors in unresolvable conditions, the age of a 
stellar population can be seriously under-estimated by the conventional SSP
model. Therefore, considering the common existence of BSS component
in real stellar populations, a considerable amount of alternations on 
the conventional ISEDs should be expected when applying the technique of
evolutionary population synthesis (EPS) to more complicated stellar systems.

\end{abstract} 


\section{INTRODUCTION}

By many years of substantial practice, evolutionary population synthesis 
(EPS) has been proved to be a powerful tool for the study of the evolution 
of stellar systems. 
EPS method is based on the great success of the understanding of stars, 
including stellar structure and evolution (without taking into account 
rotation, magnetic field and binarity), and the properties of stellar 
radiation from their atmospheres. 
As the basic building blocks of synthetic spectra of galaxies, simple 
stellar populations (SSPs) are built with the integration of the library 
of isochrones based on the 
theoretical evolutionary tracks, the library of stellar spectrum of individual
stars, and the stellar initial mass function (IMF). The first two ingredients
are given at a certain chemical composition, while the IMF is usually 
considered as universal (see Bressan et al. 1994, for a review). 

As the solid base for studying complex stellar systems, EPS method 
is very successful in many aspects.
In analyzing the stellar populations of composite systems such as
galaxies, EPS reproduces the overall spectral properties of the 
stellar components, and also derives star formation and 
chemical evolution history (Schulz et al. 2002).
In the case of simple stellar systems such as star clusters,
relying on almost perfect isochrone-fitting to the photometric
observations of the systems, the conventional SSP model can give
fairly good interpretations upon most of the evolutionary features
of star clusters, i.e., the main sequence (MS), red giant branch (RGB),
horizontal branch (HB) or red clump giants on the integrated light of
these components.
However, exceptions do exist.
It is often observed that a certain number of definite member stars have
holding positions in the color-magnitude diagrams (CMDs) that cannot
be predicted by the single-star 
evolution theory, and consequently the integrated contributions of these 
stragglers to the total radiation are ignored by the conventional SSP model.
Most of the unresolved objects are related to binary scenario and stellar 
collisional events, such as blue straggler stars (BSSs) in stellar systems.
In the case of open clusters, being the most luminous blue objects, 
BSSs can raise a big challenge to the SSP model (Deng et al. 1999;
Schiavon et al. 2004).

Since the first identification by Sandage (1953) in the  
cluster M3, BSSs have been observed in almost all star clusters, 
dwarf galaxies and presumably in all other stellar systems
(Stryker 1993). These enigmatic stars locate bluer and brighter of 
the cluster turnoff region in a CMD, and appear $\it{straggling}$ 
away from the regular evolutionary path, hence are named 'blue stragglers'. 
As for the nature of these objects, observations suggest that BSSs are 
more massive than the regular stars in the turnoff region, thus they
are most likely the remnants of stellar merger events, or still 
binary systems with luminous blue components.

The primary ways of the formation of BSSs are dynamical evolution of close 
binaries, and collisions in the high density area in the clusters. 
(Ferraro et al. 1997; Piotto et al. 1999; Ferraro et al. 2003). 
As suggested by mounting evidence, if BSSs are partially the result
of mass transfer or coalescence during resonant interactions involving 
binary systems (Leonard 1989; Bacon et al. 1996), 
they would be the visible tracers of binary populations, 
and additionally they provide an opportunity for learning 
how interactions in binary systems affect stellar evolution 
(Pols \& Marinus 1994). 
Meanwhile, appreciable frequency of single-single star collisions
generally occur only in the cores of the densest clusters 
(Hills \& Day 1976), where BSSs can provide observational constraints 
on stellar collision cross sections for tidal captures, direct 
collisions, and the corresponding productions in dense stellar 
environment (Bailyn 1995; Lombardi et al. 2002). 
Regardless of the origins of BSSs, when mergers occur, many of the
cluster properties are affected. Remarkable central concentration
of BSSs provides an effective examination on cluster dynamics, such 
as two-body relaxation in binary systems, core collapse rate, 
mass segregation and depletion of low mass stars in stellar systems, 
and energy equipartition among the cluster member stars.
Moreover, BSSs do exist in significant numbers in both young and old 
populations.  They provide a remarkable hotter component with respect to
the rest of the cluster (Manteiga et al. 1989) and bring on possible 
bluer integrated colors and integrated spectral energy distributions 
(ISEDs) (Deng et al. 1999).  
Therefore, the presence of numerous BSSs has to be accounted in 
population synthesis definitely for star clusters, 
and possibly for dwarf ellipsoidals, ellipticals and other galaxies. 
It is important to gather empirical information on BSSs in order to 
characterize correctly the overall properties of populations in 
stellar systems according to integrated light studies. 

Based on the assumption that all members are born simultaneously
therefore having the same age and metallicity, open and globular 
clusters are considered as the best SSP templates in the real 
world (Bica \& Alloin 1986; Battinelli et al. 1994). 
In this paper, our attention is put on the BSSs in Galactic open 
clusters. The reason is that, for the chemical composition range 
covered by Galactic population I clusters, BSSs are the most luminous 
blue objects whose contributions to the integrated light are more 
prominent, and this is especially outstanding for old open clusters 
when the red clump instead of blue HB is populated.
The BSS effects on the ISEDs become far less important for the Galactic
globular clusters, where the most luminous blue objects are the HB stars.

Since the direct observation of cluster integrated light 
(Bica \& Alloin 1986) can not attack the problems of field 
contamination and gravitational evaporation of low-mass stars,
it is not possible to get the proper ISED of a cluster in that way.  
Pure theoretical efforts based on the current theory of stellar 
evolution also fail at this point simply due to the stragglers. 
Keeping these in mind, we propose a semi-empirical approach of building 
cluster ISED in our previous work (Deng et al 1999): after careful 
membership analysis, the general stellar 
ingredients well fitted by a theoretical isochrone is represented by 
the conventional SSP model, in which the 'missing' low mass stars 
can be recovered with proper initial mass function, 
so that the true meaning of SSP is conserved;
and all the stragglers are included into the ISED by derivation of the
spectra from observations.

The importance of BSSs on the studies of stellar formation and evolution,
and population analysis of galaxies have been stressed by numerous
work (see Stryker 1993 for a review). What we are going to show in this
paper, by using old Galactic open clusters, is that the effects of 
BSSs on the integrated light
of a stellar system and on the theoretical model of SSP.
With a sample of twenty-seven Galactic open clusters of ages 
$\geq$ 1 Gyr, we present in this work a set of integrated spectra 
and modified integrated B-V color of the clusters. 
Comparison between the conventional approach and our results is made 
aimed at investigating BSSs contribution to the conventional SSP model.
A short report of our working sample is given in the next section. 
In Section 3, population synthesis is performed for each cluster. 
BSSs are treated as hydrogen-burning MS stars, whose spectra
are selected from Lejeune (et al. 1997,1998) stellar spectral library,
given the physical quantities by fitting evolutionary tracks to
their observed positions in the CMDs.
The other stellar ingredients in a cluster are approximated with 
Padova isochrone (Bertelli et al. 1994). The final ISED of a cluster 
is the combination of these two components. 
BSSs contribution to the conventional ISEDs varies for different
clusters depending on the basic quantities of the cluster and the
BSS component. Discussions addressing these issues are
given in Section 4. 
In Section 5, BSSs modification to the integrated properties of SSP model
in terms of wide-band B-V color is presented, and the corresponding  
consequences on age determination of a SSP is also considered.
The concluding remarks on our work are presented in Section 6.

\section{THE WORKING SAMPLE}

Photometric data of a large number of Galactic open clusters and their 
BSS members were published by Ahumada \& Lappaset (1995, AL95 hereafter). 
This data set enables good access to the study of BSS population in 
open clusters. 
Due to the reasons discussed below, only twenty-seven Galactic open 
clusters of ages $\geq$ 1Gyr are selected from AL95 as our working sample. 
For the younger clusters, there are no clearly defined turnoff point, therefore 
definite BSSs identification cannot be given, 
and due to short lifetime and usually low member star richness, 
some phases in the CMD are not populated to ensure good statistics. 

The basic parameters of the selected clusters are given in Table 1,
where columns 1-3 give the cluster name and equatorial coordinates 
R.A \& Dec. (2000); columns 4-7 are the ages, color excesses (E(B-V)), 
distance modula (DM) and metallicities (Z) of the clusters; the 
N$_{BS}$ (number of BSSs) 
and N$_2$ numbers are listed in columns 8 and 9, respectively; and 
finally, column 10 is the reference-number. 
The N$_{BS}$ numbers and BSSs photometric data for selected clusters 
are directly quoted from AL95. The N$_2$ number listed in column 9 in 
Table 1 is also from AL95, which is defined as the number of stars 
within 2 magnitudes interval below the turnoff point for a given cluster.
The other basic parameters of the selected clusters,
specifically the age, Z, E(B-V) and DM, are extracted from the more recent 
photometric and theoretical work when results newer than AL95 are available. 

N$_2$ is regarded as a very important parameter indicating the 
richness of theoretical stellar component in the simple population
synthesis scheme. For each cluster, this count is made from the same CMD 
where the BSSs are selected. Then, the ratio of N$_{BS}$/N$_2$ can be 
taken as a specific BSS frequency in a cluster, and can be used as a probe
for the cluster internal dynamic processes concerning BSS formation.
Figure 1 is the cluster number distribution of twenty-seven sample open 
clusters as functions of four different parameters: Age, Z, N$_{BS}$ 
and N$_{BS}$/N$_2$. 
There is scarcely very old clusters in our working sample. The oldest one
is 8 Gyr. More than one half of the clusters are metal-poor with respect
to solar abundance. The metallicity distribution peaks at less than a 
half solar value (Z=0.007). The cluster distributions in both N$_{BS}$ and 
N$_{BS}$/N$_2$ are inclined to low numbers, which means that the clusters
have neither large number of member stars nor BSSs. Only two clusters
possess BSSs more than thirty. The ratio of N$_{BS}$/N$_2$ is confined
in a region between 0.05 and 0.25. 
Figure 2 shows us the BSS numbers of our sample clusters as functions of 
cluster age and N$_2$ number, where there is no indication of any correlation
between N$_{BS}$ and cluster age (Fig. 2a), while a good correlation with 
N$_2$ is well defined (Fig. 2b).

It is worth emphasizing here that not all the clusters have complete 
membership determinations in terms of both proper motion and radial 
velocity. The selection of BSS candidates in AL95 is basically according 
to their appearance regions in the observed CMDs. 
In our work, all the BSSs are treated equally without any 
further detailed membership measurements and BSSs identification.  
Such a treatment inevitably suffers some problems which will be
addressed in Section 4. 

\section{THE SYNTHETIC SED OF THE CLUSTER}

An ideal SSP model could be applied to an open cluster in terms of 
metallicity and age of its members, if all stars in it were single 
(at least not interactive binaries), and the low-mass members were 
conserved against the tidal effects. 
Unfortunately, the nature turns us down at this point. 
The obvious differences between a SSP and a real star cluster should be
the stars resulted from interacting binaries, stellar collisions and
evaporation of the low mass components. 
In old open clusters, since RGB stars and red clump giants have been well 
populated, the red stragglers can hardly alter the integrated light of a
SSP due to their rare numbers and low luminosities compared with RGB and
the clump giants within the same cluster. 
Therefore, the most prominent features in the CMDs in open clusters are
the luminous BSSs. 

According to above discussions, the stellar population of a cluster is assumed 
to be composed of two components in our work. 
The first one is a package of all the member stars fitted by an isochrone 
in the CMD, plus the low-mass stars that have been peeled off 
by tidal effects. This is obviously nothing but the conventional SSP model 
and named as the SSP component for use in the following text.
The second one includes all the other members straggling away from the 
isochrone. In the light of above discussions, only BSSs are considered 
into this component. We label it as the BSS component.

Assuming that the cluster CMD can be well fitted by a theoretical model,
the SSP component of a cluster can be substituted with a theoretical 
isochrone that best fits the age and metallicity of the cluster.
While, the BSS component is specified in the CMD with photometric data of 
AL95. Figure 3 shows the composite CMD of NGC 6791 as an example, where the 
solid line is the isochrone, the dotted line is the zero age main sequence 
(ZAMS), and the dots are the BSSs defined by photometry. We notice that
there is a number of objects located below the ZAMS loci. This is partly
due to observational uncertainties.

\subsection{Getting the theoretical spectra of the BSSs}

It is a growing consensus that the major mechanisms of BSS production
are mass-transfer and merger events in close binary systems, or 
stellar collisions in dense environment (Sch\"{o}nberner \& 
Napiwotzki 1994). The physically interactive processes in binary 
systems will transfer fresh hydrogen fuel from hydrogen-rich 
envelop to hydrogen-depleted or exhausted core and re-ignite it, 
which makes the remnants behave like MS stars (Benz \& Hills 1987,1992), 
as such, BSSs are treated as hydrogen-burning MS stars in our 
work. Their masses, luminosities and effective temperatures
can be determined by
fitting their positions in the CMD with evolutionary tracks 
of higher masses than the turnoff. Given the effective temperature 
($\it{T_{eff}}$) and surface gravity ($\it{log}$ $\it{g}$), a theoretical spectrum
from the library (Lejeune et al. 1997,1998) can be assigned to the BSSs.
This process is performed for all the BSSs in our sample clusters.

As shown in Figure 3, there are some BSSs located below the ZAMS. 
Except for possible photometric errors, their missing-positions 
are most likely due to the helium-enrichment in the atmosphere after 
merger events (Benz \& Hills 1987,1992). The intrinsic properties 
of these stars need more carefully observational studies. 
In our work, these low-luminosity BSSs are not considered in 
computing the ISEDs of the clusters, since their contributions to 
the integrated light are even much lower than the stars in the turnoff region.
 
\subsection{Getting the ISED of a cluster}
The purpose of our work is to build modified ISEDs based on the observations 
of open clusters, so the clusters as a whole will be treated as unresolvable 
stellar populations. 
Most of the cluster member stars can be in principle confined in the 
single-star evolution regime, including the unresolved binaries shown in 
photometric binary sequence in the CMDs. 
As the SSP component is based on single-star evolution theory, we should 
emphasize here that the IMF applied to conventional 
SSP model should account for both the single stars and the photometric binary 
stars in the observational CMDs. 
For a given cluster of age $\it{t}$ and metallicity $\it{Z}$, assuming
Salpeter IMF $\it{\phi(m)}$, the ISED of SSP component is,
\begin{equation}
\textit{F}_{iso}(\lambda,t,Z)=\textit{A}\int_{m_l}^{m_u}
\phi(m)\textit{f}(\lambda,m,t,Z)\textit{d}m
\end{equation}
where $\it{f(\lambda,m,t,Z)}$ is the flux of a star of mass $\it{m}$, 
age $\it{t}$ and metallicity $\it{Z}$. $\it{A}$ is a normalization 
constant which is fixed by the cluster richness parameter N$_2$. 
$m_u$ and $m_l$ are the upper and lower integration limits, in the current
context, $m_u$ is the initial mass of the most massive living star at 
the age of the cluster, while $m_l$ takes its regular meaning.

By using Salpeter IMF, $\it{A}$ is derived from the formula: 
\begin{equation}
\textit{N}=\textit{A}\int_{m_1}^{m_2}m^{-2.35}\textit{d}m=\textit{N}_2
\end{equation}
where $\it{m_1}$ and $\it{m_2}$ are the initial masses corresponding to 
the two limiting magnitudes defined by N$_2$, i.e. $\it{m_2}$ is 
the initial mass of the very star in the turnoff, and $\it{m_1}$ is the 
initial mass of star whose magnitude is two magnitudes 
lower than that of the cluster turnoff point.

For the BSS component, the integrated spectra $\it{F_{BS}(\lambda,t,Z)}$ is 
given by directly summing up the individual spectrum,
\begin{equation}
\textit{F}_{BS}(\lambda,t,Z)=\sum_{i=1}^{\textit{N}_{BS}}\textit{f}_{BS}^i
\end{equation}
where $\it{f^i_{BS}}$ is the theoretical spectrum for a BSS, the index 
$\it{i}$ runs from 1 to $N_{BS}$. N$_{BS}$ is the total number of BSSs 
in a cluster.

Finally, the ISED of a cluster in our work is the combination of 
the integrated spectrum of SSP component $\it{F_{iso}(\lambda,t,Z)}$ and  
that of BSS component $\it{F_{BS}(\lambda,t,Z)}$. 

\section{BSS MODIFICATION TO THE CONVENTIONAL ISED}

As discussed in above text, there are several channels of BSSs formation.
Depending on the specific dynamical environment 
of the host clusters, and the correlation with the richness of the cluster
(see Fig. 2b), BSS population in different clusters actually presents 
very different properties. Therefore, BSSs effects on the final ISEDs 
also vary greatly among the clusters in our sample. 

To demonstrate the effects of BSSs on the conventional ISEDs in the case 
of open clusters, Figure 4 is given with the fluxes of different ingredients, 
where the abscissa is the wavelength in angstrom, the integrated spectrum of 
the BSS component in dotted line, that of SSP component in dash line, and the 
total ISED (combination of the SSP and BSS components) is in solid line. 
In our work, the BSS modifications to the conventional SSP model can be
divided into three cases: Normal star dominates (Fig. 4a), 
BSSs important (Fig. 4b) and BSS dominates (Fig. 4c). 
Regardless of the specific cases, the existence of BSS population modifies
all the integrated light of our sample clusters at some extent, 
and these alternations are inclined 
towards UV and blue bands, therefore turns the spectra hotter. 
For different cases, the BSS population possesses certain properties, 
i.e., locations in the cluster, membership measurements and number richness, 
which are all important in interpreting the strength of BSS effects 
on the final ISEDs. These situations are discussed separately below.

\subsection{The normal star dominate case} 

In our work, six of the total twenty-seven sample clusters, namely 
Berkeley 19, NGC 752, NGC 2243, NGC 2420, NGC 2506 and NGC 6208, are 
classified as this category.
In this case, BSSs generally locate near the cluster turnoff point 
in the CMDs, and N$_{BS}$ is a minority when compared with the turnoff 
region stars, so that they can not produce obvious contribution to 
the integrated light of the cluster (Fig. 4a). 
Based on the physical parameters and photometric data of sample clusters 
and BSSs in different work, specifications of the clusters are given 
below.

\subsubsection{NGC 6208}

The membership probabilities of the five BSSs (as in AL95) in NGC 6208 
have been an ambiguous issue.
Three of the five BSSs were classified as B stars and non-members by 
Lindoff (1972). The other two high probability member BSSs possess   
very low luminosities and locate below the ZAMS in the composite CMD. 
According to above discussion, these two BSSs can be ignored when 
building the ISED. After excluding the non-member BSSs, the ISED of 
this cluster virtually suffers no BSS correction and a conventional 
SSP model is retained.
In Figure 5, the solid triangles are the five BSSs in NGC 6208. 
Ones with open boxes are the non-member BSSs in Lindoff (1972).

\subsubsection{NGC 2420 and NGC 2506}

These two clusters, NGC 2420 ($\it{l}$=$198.1^\circ$, $\it{b}$=$+19.6^\circ$)
and NGC 2506 ($\it{l}$=$230.6^\circ$, $\it{b}$=$+9.9^\circ$), have very
similar physical parameters, the richnesses defined by N$_2$ for them are
pretty close: 140 for NGC 2420 and 130 for NGC 2506.
Although both have N$_{BS}$ numbers of twelve, 
the properties of BSS populations are quite different. 
In AL95, the BSSs data of NGC 2420 were selected from the photometric work 
of West (1967), covering a sky region of 7.5 arcmin around the cluster center. 
Four BSSs were classified as field stars in that work. 
In contrast to that, all the BSSs in NGC 2506 (McClure et al. 1981) have  
membership probabilities p $\geq$ 90\%, based on the proper motion study in 
Chiu \& van Altena (1981). 

Figure 6 shows a comparison of BSSs distribution and ISED 
modification in these two clusters. 
In the CMD in left panel, the solid circles are BSSs in NGC 2506, solid 
triangles are BSSs in NGC 2420, and solid-line isochrone represents general 
stellar components in two clusters, according to their similar age (3.4 Gyr) 
and metallicity (Z=0.007). 
Although the BSS population in NGC 2420 is generally brighter than that in 
NGC 2506, the larger N$_2$ number balances this difference and makes 
the final ISED of NGC 2420 similar to that of NGC 2506, as shown in the 
right panel in Figure 6. 

\subsection{The BSSs important case}
Over one half of our sample clusters (Berkeley 32, Berkeley 39, 
Berkeley 42, IC 4651, King 2, King 11, Melotte 66, NGC 188, NGC 1193, 
NGC 2112, NGC 2660, NGC 2682, NGC 3680, NGC 6791, NGC 6819, NGC 6939, 
NGC 7142 and NGC 7789) belong to the second case: BSS component 
seriously modifies the cluster ISED (see Fig. 4b). This situation 
generally happens when: (a). there is a large number of bright 
BSSs in the cluster; and/or (b). The N$_{BS}$ is small, but there is 
a few (sometimes even just 1!) BSSs which are much brighter than the cluster 
turnoff point; and/or (c). the cluster is too much under-populated.

As mentioned in above context, BSSs contribution directly 
depends on the selection of these bright stars. More clear
identification from high-definite photometric work certainly 
favors more accurate results.
Metallicity is also a sensitive parameter to the BSSs contribution: 
low-metallicity ISED obviously presents stronger flux at shorter 
wavelength when compared with high-metallicity ISED 
(Schulz et al. 2002).
Discussions about these problems in some sample clusters in this 
case are present in the following.

\subsubsection{King 2}
Thirty stars were cited as BSSs in King 2 in AL95, based on the 
Johnson-Cousins UBVR CCD photometry of Aparicio et al. (1990),
in which only BSS candidates within the core radius are considered,
in order to avoid the probable contamination by field stars. 
As a cluster of large N$_{BS}$ number (as shown in Figure 7), 
the effects of BSSs on the ISED of this cluster are obviously due to 
the large number of bright BSSs.

\subsubsection{NGC 6939}
As shown in Figure 8, 
four stars were defined as BSSs in NGC 6939 in Cannon \& Lloyd (1969). 
Star 59 ('59' is the identification number of the star marked in the
finding chart in the photometric work of Geisler (1988). All the numbers
in the following are the same identifications) was mentioned as a BSS 
in Geisler (1988), although it was unlikely a member. 
In AL95, NGC 6939 has N$_2$ number of eighty which is
smaller than that of King 2 and bigger than that
of NGC3680, thus NGC 6939 has an intermediate richness in our working sample. 
In this situation, the effects of BSSs on the ISED should be better 
attributed to the high luminosities of the BSSs, since the N$_{BS}$ of 
this cluster is much smaller than that of King 2.

\subsubsection{Berkeley 42 and NGC 3680} 
These two clusters possess small numbers of both N$_{BS}$ and N$_2$.  
The single BSS in Berkeley 42 was selected from the CCD photometry of 
Aparicio et al. (1991). 
Four stars (29, 33, 43 and 56) were identified as BSSs in NGC 3680 in AL95, 
according to the photo-electric observations of Eggen (1969), who 
later (Eggen 1983) identified stars 3, 29, 33 and 56 as BSSs and stars 
20 and 34 as 'red stragglers' with ubvy photometry.
In the CCD photometry of Anthony-Twarog et al. (1991), stars 29 and 33 
were considered as probable cluster members, but star 56 is unlikely a member. 
Meanwhile, an important presence of binaries around the turnoff point of 
NGC 3680 was observed by Anthony-Twarog et al. (1991). 
As shown in Figure 9, there is a BSS in both two clusters whose 
positions in the CMDs highly exceed the cluster turnoff that contributes 
most of the light in the blue part of the ISED. 
 
\subsubsection{Melotte 66}
Melotte 66 holds a really special position in our work. Whether it 
should be included in this case or not is still not certain.
Based on the photoelectric and photographic photometry in Hawarden 
(1976), the BSSs in Melotte 66 were selected in the cluster central region 
of 7 arcmin. However, 
the BSSs ISED plotted as dotted line in Figure 10 is not at all 
bluer than the conventional SSP spectrum, 
therefore the B-V color of the cluster is not modified (shown in Table 2), 
even though it has the largest N$_{BS}$ 
of forty-six among all the sample clusters.

Furthermore, we find that forty BSSs of the cluster have rather red 
spectra instead of blue ones compared with the turnoff. The remaining 
six blue-spectrum BSSs are all located in 
Ring I of 3 arcmin (Hawarden 1976). In Figure 11, solid circles are the 
blue-spectrum BSSs. The solid rectangles are red-spectrum BSSs which 
should be better classified as 'yellow stragglers'. 
This complicated situation can be partially attributed to the 
significant metallicity dispersion in and around the turnoff region 
of the cluster (Twarog \& Anthony-Twarog 1995), apart from the membership
and observational uncertainty issues. 

\subsubsection{NGC 188}

As one of the most studied old open clusters, 
NGC 188 has been the subject of numerous studies in many aspects, 
certainly including its BSS population.
All the twenty BSSs in AL95 were selected from Ring I and 
II in the finding chart of Sandage (1962), with membership probabilities 
p $\geq$ 80\% (Dinescu et al. 1996). This BSS number almost doubles 
the number of eleven from an earlier work of Eggen \& Sandage (1969).
All the BSSs show obvious high central concentration. 

In the work of Dinescu et al. (1996), nine out of eleven of the BSSs in 
Eggen \& Sandage (1969) were confirmed as cluster members with high 
probabilities, while the other two (D and I-102 in Sandage 1962) were 
excluded due to their 0\% proper motion membership probabilities.  
Dinescu (et al. 1996) selected eleven probable BSSs with
$P_{\mu}$ $\ge$ 50\% and $P_{\mu,r}$ $\ge$ 70\% ($P_{\mu,r}$ is 
proper motion membership
probability, and $P_{\mu,r}$ is the combined probabilities of both proper
motion and spatial distribution), which include five of Eggen \& Sandage (1969)
while adding six new BSSs. Among the six new BSSs, four are out of Ring-III and
the other two are in Ring-II of Sandage (1962). Although being the latest
observation of the cluster, Dinescu's BSS catalog 
does not show central concentration significantly higher than RGBs, therefore
the original AL95 catalog is still adopted in our work.

The BSS population in this cluster has been followed intensively in the
past. Although a few of BSSs are subject to dispute on their membership
probabilities, the overall BSS population still forms a considerable 
amount of modification to the total ISED of the cluster. This correction 
to the corresponding SSP model at the age and metallicity of NGC 188 is 
very convincing based on so much both theoretical and observational
efforts.

\subsubsection{NGC 2682}

Similar to NGC 188, NGC 2682 (M67) has also been regarded as a classical
old open cluster due to so much research work on almost all the subjects 
in stellar physics.
The photometric data of its BSSs in AL95 was from 
Sanders (1989) who derived the membership measurements for the
cluster stars at a earlier time (Sanders 1977). 
Among the total thirty BSSs in AL95, twenty-two were confirmed and two
more were added by Deng et al.(1999), based on the photometric data of 
Beijing-Arizona-Taipei-Connecticut (BATC) sky survey and membership 
data from Girard et al. (1989).
Figure 12 is given here to compare the possible discrimination between 
these two groups of BSSs identification. 
In the CMD (left panel), the solid dots are BSSs in AL95, open rectangles 
are BSSs defined by Deng et al. (1999). As shown in right panel, there 
is no obvious difference in the ISEDs of these two BSSs identification. 
The result confirms our previous work of BSSs modification 
to the SSP model in this cluster (Deng et al. 1999).

\subsection{The BSSs dominate case}

For the third case (Tombaugh 1, NGC 1252 and NGC 2158), there is at least 
one extremely luminous BSS in the cluster and/or the cluster is extremely 
under-populated. The BSS(s) is so bright that controls almost 
all the integrated light of the cluster (Fig. 4c). The situation in this 
case raises issues such as membership probability of the bright star, 
and the property of such star if it is a member, just as the luminous 
BSS F81 in NGC 2682 (Deng et al 1999).

\subsubsection{Tombaugh 1 and NGC 1252}

AL95 gives only one BSS for Tombaugh 1 (Turner 1983). It appears that 
the cluster is a faint system that is merely visible from the field 
background. Turner (1983) obtained photoelectric photometry for only 
twenty-six stars in the cluster region.
In the finding chart of Turner (1983), the  
BSS locates leaning on but outside the Ring I of a radius of 5 arcmin. 
In the CMD of Tombaugh 1 (left panel in Figure 13), it lies more than one 
magnitude higher of the cluster turnoff point. 
In the ISED (right panel in Figure 13), this single BSS contributes about 
70\% of the total ISED. Especially in the wavelength shorter than 3000 
${\AA}$, it provides almost all the integrated light of the cluster.
In this instance, the BSS is not much brighter than the turnoff, the 
overwhelming effect of the BSS is completely due to the under-abundant 
cluster members.
Meanwhile, a B9.5 V star is identified as the only confirmed member BSS in 
NGC 1252, based on the photometric study by Bouchet \& The (1983).
It suffers the same problem as that on Tombaugh 1 on having 
very small numbers of both N$_{BS}$ and N$_2$ (see Table 1). 
We argue that in this case, more detailed photometric work and
membership measurements are inevitably needed for BSSs identification. 
For the moment, the ISED derived in this case is considered as direct 
superficial exercise, whose implication on the conventional SSP model 
corresponding to the age and metallicity of the cluster has little 
statistical significance.

\subsubsection{NGC 2158}

Based on the photo-electric photometry of Arp \& Cuffey (1962), thirty
stars were cited as BSSs in NGC 2158 in AL95. 
Among them, twenty-two are located in the central region of the cluster 
(Ring I and II), while the other eight scattered in Ring III and IV. 
Four BSSs were confirmed as double stars by Arp \& Cuffey (1962), 
including the brightest BSS of the cluster (locate in Ring II), 
which is marked as 'D1' in Figure 14.
D1 is more than 4 magnitudes brighter than the cluster turnoff point. 
This magnitude gap exceeds the maximum limit for BSS predicted by the 
mass transfer theory (McCrea 1964), just like F81 for M67 (Wheeler 1979). 
A new model is needed to interpret such a BSS location in the CMD, if
such a star is a real member of the cluster.

To show the relative alternation of these bright stars to the 
conventional SSP model, Figure 15 is given here to show the flux 
of BSSs in different combinations in NGC 2158. The thick solid line is 
conventional ISED corresponding to the isochrone only. The other four 
lines are respectively different BSSs combinations: the long dash line 
is the isochrone plus all BSSs except D1; the short dash line is the 
spectrum of D1 alone; the dotted line is for all BSSs contribution; 
and the thin solid line is isochrone plus all BSSs. 
As shown in Figure 15, D1 dominates the ISED from UV to near IR 
wavelengths. It contributes almost all the energy at the spectrum 
shorter than 1500 ${\AA}$, about 80\% of the integrated light of all 
BSSs until 4000 ${\AA}$, and still around 50\% up to 7000 ${\AA}$ 
and beyond.

\section{Modifications to ISED in terms of wide-band B-V color}

Though the effects of BSSs on SSPs' ISEDs have been clearly demonstrated 
by the spectral alternations of composite clusters, wide-band color is 
in practice more frequently used when trying to understand unresolved stellar
populations in galaxies. 
Compared with the conventional SSP model, BSS turns the spectrum bluer. 
The photometric bluer color in nature corresponds to the younger stellar 
populations, in other words, the age will be under-estimated when fitting 
the color with conventional SSP model.
In terms of the modification to wide-band B-V color, BSS correction on 
the SSP model can be quantified.

\subsection{B-V color versus Age}
The integrated color of a theoretical SSP changes with time when massive 
stars leave the blue and luminous MS. Here, the most intuitionistic trend in 
the picture of conventional SSP model is that the color becomes redder. 
However, the common existence of BSSs will put a different signature on this 
scenario. In our work, the 'observed' B-V color with BSSs effects  
is quite bluer than that predicted by the conventional SSP model.
Figure 16 depicts such a modification. 
Theoretical B-V color of four different metallicities is given in four lines
in different types, obtained by convolving the conventional ISEDs made with 
standard SSP model (assuming Salpeter IMF and using Padova stellar track library) 
with corresponding filter responses. 
The solid circles are B-V color (derived with the combined ISEDs)
of the twenty-seven sample clusters involving BSSs contribution. 
It is clearly shown that the B-V color has been dramatically modified by BSSs. 
All the solid circles lie below the theoretical value of Z=0.02, although 
40\% of the sample clusters are metal-rich (Z $>$ 0.02). 
Three lowest points in Figure 16 are those BSS-dominating clusters. 
As discussed in Section 4, quantitative analysis of BSSs effects on this 
kind of clusters should be supported by further studies both in observation
and theory. 
Detailed results are listed in Table 2. Columns 1-3 give the name, age, and 
metallicity of the cluster (age and metallicity parameters follow the same 
sources as in Table 1). 
Column 4 is N$_2$ number. Column 5 is the original 
BSSs number listed in AL95, while column 6 is the adopted BSSs number in our 
work (eliminating the BSSs below ZAMS). Finally, columns 7-8 are B-V color 
resulted from the conventional SSP model and from the BSS-corrected model, 
respectively.

Assuming that all the sample clusters have solar metallicity (Z=0.02), 
Figure 17 shows quantitatively the age underestimation when fitting the 
photometric data with conventional SSP model.
The solid line in Figure 17 is theoretical B-V color of Z=0.02 against 
the age of our sample clusters. Keeping B-V color while doubling the age 
we get the dash line, and triple the age we get the dotted line. 
In Figure 17, the smallest deviated points in color can be approximated with 
30\% age-expanding conventional SSP model (long-dash line), and the 
double-age dash line runs through many of the BSS-corrected solid circles. 

\subsection{B-V color versus Metallicity}

As described in Section 2, there is a non-negligible metallicity 
dispersion in our sample clusters. More than a half of them are 
metal-poor (Z $<$ 0.02).
Since metallicity is an essential parameter for evolution of both
stars and galaxies, it will inevitably influence the intrinsic
properties of BSSs, therefore brings changes on the BSSs effects on
conventional SSP model. 

In this work, the possible impact of metallicity on BSSs contribution
is detected with the corrected B-V colors as a function of cluster
metallicities. Corresponding details are pointed out in Figure 18, where
the ordinate is the ratio of B-V alternations caused by BSS component
in a cluster to the theoretical B-V color of conventional SSP model, or
named relative alternations, and the abscissa is the metallicity of our
sample clusters. 
It seems from Figure 18 that cluster metallicity and BSS contribution 
are well correlated in the metal-poor case. This is simply stressed 
in Figure 19 with four typical Z values (Z $\leq$ 0.02), each of 
which has not less than three samples in this work. In Figure 19, 
the relative alternations in B-V color increases with metallicity.
However, such a correlation does not show in the case of 
Z $>$ 0.02. We need a complete working sample for this analysis. 

\subsection{Fitting the ISEDs of open clusters with conventional 
SSP model: uncertainties}

Based on the expatiation in above paragraphs, conventional SSP model 
derived from the single star evolution theory has been dramatically 
altered by BSS component in a population. This modification
typically enhances the UV and blue parts of the ISEDs, 
therefore the B-V color of SSPs 
becomes bluer. If it is to derive an age to a population including
BSSs by fitting a conventional SSP model, a substantial uncertainty
will be there. Such a conclusion can be visualized by
direct fitting the real ISED of a population with the
conventional SSP model with depressed either age or 
metallicity, since both factors make the ISED bluer as the BSS components
do as in the case of our sample clusters.
In other words, ignoring the existence of BSSs in stellar populations
of metallicity and age ranges covered by our sample may seriously
under-estimate either age or metallicity of the population by the
conventional SSP model.

In this work, the composite ISEDs of open clusters 
including BSSs contribution are fitted with the conventional SSP model
at younger ages or lower metallicities. A quantitative measure
of the uncertainties introduced by BSSs in a population can be
given in this way.

Taking the most popular old open clusters NGC 188 and NGC 2682
as examples, the results of best fitting are presented 
in Figures 20 and 21, respectively. The abscissa is the wavelength in 
angstrom. The ordinate is the logarithmic value of the absolute flux
of ISED normalized at wavelength of 5500 ${\AA}$.
In the lower parts in figures, the difference between the composite
ISED of the cluster and the best-fitting conventional SSP ISED 
is given as $\delta$, together with the standard 
deviation $\sigma$ in the region of $\pm$3$\sigma$ in dotted lines.
In each figure, the left panel is the fitted result that keeps 
the same metallicity but younger age, while the right panel is that 
just the opposite, keeping the same age but lower metallicity. 
The composite ISED of the cluster is plotted in solid line, the 
conventional ISED is given in dash line. As a comparison, the basic 
parameters of the cluster adopted in our work are given in 
the top right corner in each plot. The parameters of the fitted conventional 
ISED is given just below the fitting line. 

As shown in Figures 20 and 21, conventional SSP model can fit 
perfectly most of the features of the composite ISED of the cluster, 
if we let the
age or metallicity free. The exception is at the extreme UV part of
the SED. As we can see from the best fitting parameters, both age and
metallicity can change by a factor of 2 in our examples.

\section{SUMMARY AND DISCUSSIONS}

Aimed at detecting BSSs effects on the integrated properties of conventional 
SSP model, twenty-seven Galactic open clusters older than 1Gyr and
their BSS contents are studied in this work. 
In general, the ISEDs of our sample clusters have been dramatically 
modified, and such modification is emphasized in UV and blue bands; 
In terms of photometry, the integrated B-V color of the sample clusters 
becomes bluer, this leads to an under-estimation either in age or  
in metallicity when fitting the observed colors 
of a target unresolved population 
with that of a conventional SSP model. The present results are summarized
and discussed in the following, 

\begin{enumerate} 

\item In this work, twenty-seven Galactic open clusters older than 
1Gyr are selected as our working sample. The photometric data of BSSs 
including B-V colors and V-band magnitudes, N$_{BS}$ and N$_2$ number are 
directly quoted from AL95. The basic parameters of sample clusters from 
recent literatures are used when they are available. 

\item We scheme a semi-empirical approach of building cluster ISED. 
The general stellar ingredients including all member stars well 
fitted by a theoretical isochrone are represented by the conventional 
SSP model. 
BSSs are treated as hydrogen-burning MS stars. Their positions in the 
CMD are fitted with evolutionary tracks of masses higher than the 
turnoff. Then, a theoretical stellar spectrum is assigned to each BSS, 
according to parameters ($\it{T_{eff}}$ and $\it{log}$ $\it{g}$) derived 
from the best fitting. Finally, the combination of these 
two components constructs the modified ISED of the cluster. 

\item We have divided the BSSs modifications to conventional 
ISEDs into three cases, namely, normal star dominate (Fig. 4a), 
BSSs important (Fig. 4b) and BSSs dominate (Fig. 4c). 
The modifications are mainly related to the BSSs richness in the 
clusters and their positions in the observed CMDs. 
For the first case, BSSs generally locate near the cluster turnoff 
in the CMD, and the number of BSSs is small when compared with 
turnoff region stars, therefore they can not produce obvious 
contribution to the ISED of the cluster. 
For the second case, BSSs effects get enhanced generally due to: 
(a). there is a large number of bright BSSs in the cluster; 
(b). The N$_{BS}$ is small, but there is a few (even just 1!) 
BSSs which are much brighter than the cluster turnoff point; 
and (c). the cluster is very member-poor. 
The third case is an excessive situation. There is at least one 
extremely luminous BSS in the cluster and/or the cluster is extremely 
poorly populated. The bright star(s) is so bright that controls almost 
all the integrated light of the cluster. 

\item We have depicted the alternation of B-V color of our sample 
clusters caused by BSSs. As shown in Figure 16, the B-V color 
becomes significantly bluer. 

\item We have discussed the effects of metallicity on BSSs contribution. 
As shown in Figure 18, they are well correlated in 
the metal-poor case (Z $<$ 0.02). In Figure 19, the ratio of the 
alternation of B-V color increases with metallicity increasing. 
But the correlation does not follow the case of Z $>$ 0.02. 

\item We have scaled the age underestimation of a cluster by fitting 
the observed B-V color with conventional SSP model. Based on the 
assumption that all the sample clusters have solar metallicity (Z=0.02), 
we have shown in Figure 17 that the age of an observed stellar 
population is seriously under-estimated by conventional SSP model. 
As clearly shown in Figure 17, 30\% uncertainties in ages can be given
for the least affected clusters, and for many cases, such corrections
amounts to twice of the conventional age.

\item We have also detected the age and metallicity uncertainties when 
fitting the observed cluster ISED with conventional SSP model, since 
decreasing either age or metallicity will also make the ISED bluer as
inclusion of BSSs does. 
Taking old open cluster NGC 188 and NGC 2682 as examples, the 
results of the best fitting show that conventional SSP model can fit 
perfectly most of the features of the composite ISEDs including BSSs 
contribution if we let the age or metallicity free, and both age and 
metallicity can change by a factor of 2. 

\item The modification due to BSSs to the theoretical SSP model 
corresponding to each cluster show that BSS population is 
very important, and can be crucial for stellar population analysis of 
complex stellar system. 
This is especially true when applying the current EPS
technique to unresolved stellar systems. 
Considering open clusters as typical descendants of regular star forming
activities, and regulating our results to the age and metallicity ranges
covered by present data set, SSP model derived from single star evolution 
theory is seriously altered. Therefore, the contribution of such a component 
should be considered when applying EPS method to more complicated stellar 
systems.

\end{enumerate}

The present results rely on the photometry from different sources, and the 
formation mechanisms and the specific properties of BSSs in each cluster 
vary greatly among the sample, therefore we are limited to draw our 
conclusion on a single cluster base. With careful cluster physical 
parameter measurements and membership analysis of individual BSSs, the 
current results are still meaningful and can be regarded as reference for 
further investigation on this problem. To reveal the effects of BSSs on
stellar populations in all different cases of age, metallicity and
environment, systematic observations together with simulations of stellar
dynamical processes at all conditions are needed.
 
\acknowledgements
We would like to thank the National Science Foundation of China (NSFC)
for support through grants 10173013 and 10273021, and the Ministry of 
Science and Technology of China through grant G19990754.

\begin{deluxetable}{llrlllllll}
\tabletypesize{\scriptsize}
\tablecaption{Parameters of the Sample Clusters \label{tbl-1}}
\tablewidth{0pt}
\tablehead{
\colhead{ID name} & \colhead{R.A.(2000.0)} & \colhead{Decl(2000.0)} &
\colhead{Age(Gyr)} & \colhead{E(B-V)} & \colhead{DM} & \colhead{Z} &
\colhead{N$_{BS}$} & \colhead{N$_{2}$} & \colhead{Refe.}
}
\startdata
Tombaugh 1 &07 00 29 &-20 34 00  &1.0        &0.40        & 13.60        &0.02         & 1 &  5 &(1) \\
NGC 6208   &16 49 28 &-53 43 42  &1.0$^2$    &0.18$^2$    & 10.55$^3$    &0.023$^4$    & 5 & 60 &(2) (3) (4)\\
NGC 2660   &08 42 38 &-47 12 00  &1.2        &0.36        & 12.30        &0.02         &18 &110 &(5) \\
NGC 2158   &06 07 25 &+24 05 48  &1.2        &0.55        & 15.10        &0.006        &30 &150 &(6) \\
NGC 6939   &20 31 30 &+60 39 42  &1.6        &0.33        & 12.27        &0.02         & 4 & 80 &(7) \\
NGC 3680   &11 25 38 &-43 14 36  &1.6        &0.075       & 10.20        &0.026        & 4 & 18 &(8) \\
NGC 752    &01 57 41 &+37 47 06  &1.7        &0.035       &  8.25        &0.014        & 1 & 25 &(9) \\
NGC 2112   &05 53 45 &+00 24 36  &2.0$^{10}$ &0.63$^{10}$ &  9.65$^{10}$ &0.014$^{11}$ &15 & 80 &(10) (11) \\
NGC 7789   &23 57 24 &+56 42 30  &2.0        &0.24        & 12.00        &0.016        &25 &130 &(12) \\
IC 4651    &17 24 49 &-49 56 00  &2.4        &0.086       & 10.10        &0.035        & 8 & 35 &(13) \\
NGC 6819   &19 41 18 &+40 11 12  &2.5$^{14}$ &0.16$^{14}$ & 12.35$^{14}$ &0.025$^{15}$ &33 &270 &(14) (15) \\
Berkeley 19&05 24 06 &+29 36 00  &3.0$^{16}$ &0.40$^{16}$ & 13.50$^{16}$ &0.008$^{17}$ & 1 & 10 &(16) (17) \\
NGC 1252   &03 10 49 &-57 46 00  &3.0        &0.02        &  9.04        &0.02         & 1 &  7 &(18) \\
NGC 2420   &07 38 23 &+21 34 24  &3.4        &0.05        & 11.95        &0.007        &12 &140 &(19) \\
NGC 2506   &08 00 01 &-10 46 12  &3.4        &0.05        & 12.20        &0.007        &12 &130 &(20)  \\
NGC 2682   &08 51 18 &+11 48 00  &4.0        &0.045       &  9.38        &0.02         &30 &200 &(21) \\
NGC 7142   &21 45 09 &+65 46 30  &4.5$^{22}$ &0.35$^{22}$ & 11.40$^{22}$ &0.016$^{23}$ &23 &120 &(22) (23) \\
Melotte 66 &07 26 23 &-47 40 00  &4.5$^{24}$ &0.16$^{24}$ & 13.75$^{24}$ &0.008$^{25}$ &46 &180 &(24) (25) \\
King 11    &23 47 48 &+68 38 00  &5.0$^{26}$ &1.27$^{27}$ & 15.30$^{26}$ &0.012$^{27}$ &24 &140 &(26) (27) \\
NGC 2243   &06 29 34 &-31 17 00  &5.0        &0.06        & 13.05        &0.007        & 7 &120 &(28)  \\
King 2     &00 51 00 &+58 11 00  &6.0$^{29}$ &0.32$^{29}$ & 15.20$^{26}$ &0.01$^{26}$  &30 &250 &(29) (26) \\
Berkeley 42&19 05 06 &+01 53 00  &6.0        &0.64        & 10.30        &0.012        & 1 & 20 &(30)  \\
Berkeley 32&06 58 06 &+06 26 00  &6.3        &0.08        & 12.60        &0.012        &19 &150 &(31)  \\
NGC 188    &00 47 28 &+85 15 18  &7.0$^{32}$ &0.09$^{32}$ & 11.44$^{32}$ &0.024$^{33}$ &20 &170 &(32) (33) \\
NGC 6791   &19 20 53 &+37 46 18  &7.2        &0.17        & 13.52        &0.039        &27 &110 &(34) \\
NGC 1193   &03 05 32 &+44 23 00  &8.0        &0.12        & 13.80        &0.01         &16 &190 &(35) \\
Berkeley 39&07 46 42 &-04 36 00  &8.0        &0.12        & 13.40        &0.01         &29 &220 &(36) \\
\enddata

\tablerefs{
(1): Carraro \& Patat 1995,
(2): Paunzen \& Maitzen 2001,
(3): Lindoff 1972,
(4): Loktin et al. 1994,
(5): Frandsen et al. 1989,
(6): Piersimoni et al. 1993,
(7): Rosvick \& Balam 2002,
(8): Kozhurina-Platais et al. 1997,
(9): Daniel et al. 1994,
(10): Carraro et al. 2002,
(11): Brown et al. 1996,
(12): Friel \& Janes 1993,
(13): Anthony-Twarog et al. 1988,
(14): Rosvick \& Vandenberg 1998,
(15): Bragaglia et al. 2001,
(16): Christian 1980,
(17): Christian 1984,
(18): Pavani et al. 2001,
(19): Anthony-Twarog et al. 1990,
(20): McClure et al. 1981,
(21): Boyle et al. 1998,
(22): Crinklaw \& Talbert 1991,
(23): Canterna et al. 1986,
(24): Twarog et al. 1995,
(25): Gratton \& Contarini 1994,
(26): Kaluzny 1989,
(27): Dias et al. 2002,
(28): Bergbusch et al. 1991,
(29): Aparicio et al. 1990,
(30): Aparicio et al. 1991,
(31): Richtler \& Sagar 2001,
(32): Sarajedini et al. 1999,
(33): Worthey \& Jowett 2003,
(34): Kaluzny \& Rucinski 1995,
(35): Kaluzny 1988,
(36): Kaluzny \& Richtler 1989
}
\end{deluxetable}

\clearpage

\begin{deluxetable}{lcllllll}
\tabletypesize{\scriptsize}
\tablecaption{Color Modifications of the Clusters \label{tbl-2}}
\tablewidth{0pt}
\tablehead{
\colhead{ID name} & \colhead{Age(Gyr)} & \colhead{Z} & \colhead{N$_2$} &
\colhead{N$_{BS}$$^O$} & \colhead{N$_{BS}$$^A$} &
\colhead{(B-V)$_{iso}$} & \colhead{(B-V)$_{iso+BSs}$}
}

\startdata
Tombaugh 1   & 1.0  &  0.02   &   5  &  1  &  1  & 0.285  &  0.615  \\
NGC 6208     & 1.0  &  0.023  &  60  &  5  &  3  & 0.596  &  0.626  \\
NGC 2158     & 1.2  &  0.006  & 150  & 30  & 30  & 0.333  &  0.519  \\
NGC 2660     & 1.2  &  0.02   & 110  & 18  & 18  & 0.460  &  0.644  \\
NGC 6939     & 1.6  &  0.02   &  80  &  4  &  4  & 0.474  &  0.743  \\
NGC 3680     & 1.6  &  0.026  &  18  &  4  &  4  & 0.608  &  0.752  \\
NGC 752      & 1.7  &  0.014  &  25  &  1  &  1  & 0.671  &  0.713  \\
NGC 2112     & 2.0  &  0.014  &  80  & 15  & 14  & 0.543  &  0.756  \\
NGC 7789     & 2.0  &  0.016  & 130  & 25  & 24  & 0.487  &  0.763  \\
IC 4651      & 2.4  &  0.035  &  35  &  8  &  7  & 0.646  &  0.856  \\
NGC 6819     & 2.5  &  0.025  & 270  & 33  & 32  & 0.678  &  0.835  \\
Berkeley 19  & 3.0  &  0.008  &  10  &  1  &  1  & 0.790  &  0.795  \\
NGC 1252     & 3.0  &  0.02   &   7  &  1  &  1  & 0.263  &  0.849  \\
NGC 2420     & 3.4  &  0.007  & 140  & 12  & 12  & 0.764  &  0.771  \\
NGC 2506     & 3.4  &  0.007  & 130  & 12  & 11  & 0.769  &  0.771  \\
NGC 2682     & 4.0  &  0.02   & 200  & 30  & 21  & 0.762  &  0.914  \\
Melotte 66   & 4.5  &  0.008  & 180  & 46  & 46  & 0.823  &  0.823  \\
NGC 7142     & 4.5  &  0.016  & 120  & 23  & 22  & 0.634  &  0.879  \\
NGC 2243     & 5.0  &  0.007  & 120  &  7  &  7  & 0.802  &  0.819  \\
King 11      & 5.0  &  0.012  & 140  & 24  & 12  & 0.683  &  0.858  \\
King 2       & 6.0  &  0.01   & 250  & 30  & 30  & 0.681  &  0.858  \\
Berkeley 42  & 6.0  &  0.012  &  20  &  1  &  1  & 0.645  &  0.872  \\
Berkeley 32  & 6.3  &  0.012  & 150  & 19  & 18  & 0.645  &  0.873  \\
NGC 188      & 7.0  &  0.024  & 170  & 20  & 17  & 0.841  &  0.960  \\
NGC 6791     & 7.2  &  0.039  & 110  & 27  & 18  & 0.799  &  1.028  \\
NGC 1193     & 8.0  &  0.01   & 190  & 16  & 15  & 0.777  &  0.896  \\
Berkeley 39  & 8.0  &  0.01   & 220  & 29  & 28  & 0.749  &  0.896  \\
\enddata
\tablenotetext{O}{The original BSSs number listed in AL95}
\tablenotetext{A}{The BSSs number adopted in our work, eliminating the BSSs below ZAMS}

\end{deluxetable}

\clearpage

\begin{figure}
\plotone{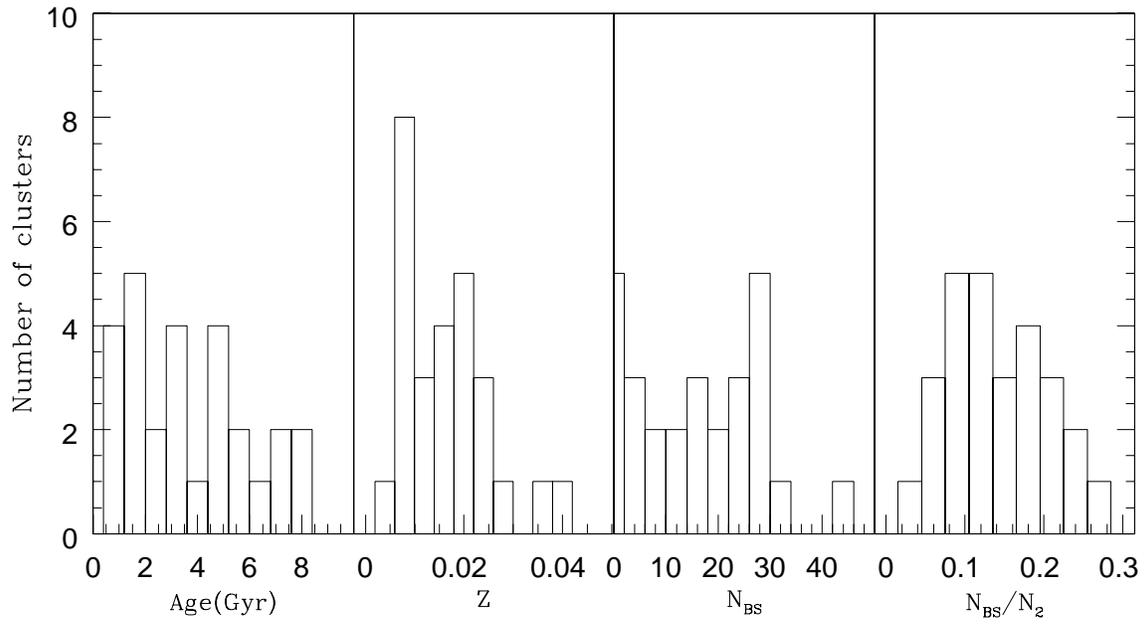}
\caption{Distributions of the open cluster sample
according to four different 
parameters: Age, Z, N$_{BS}$ and N$_{BS}$/N$_2$.\label{fig1}}
\end{figure}

\begin{figure}
\plotone{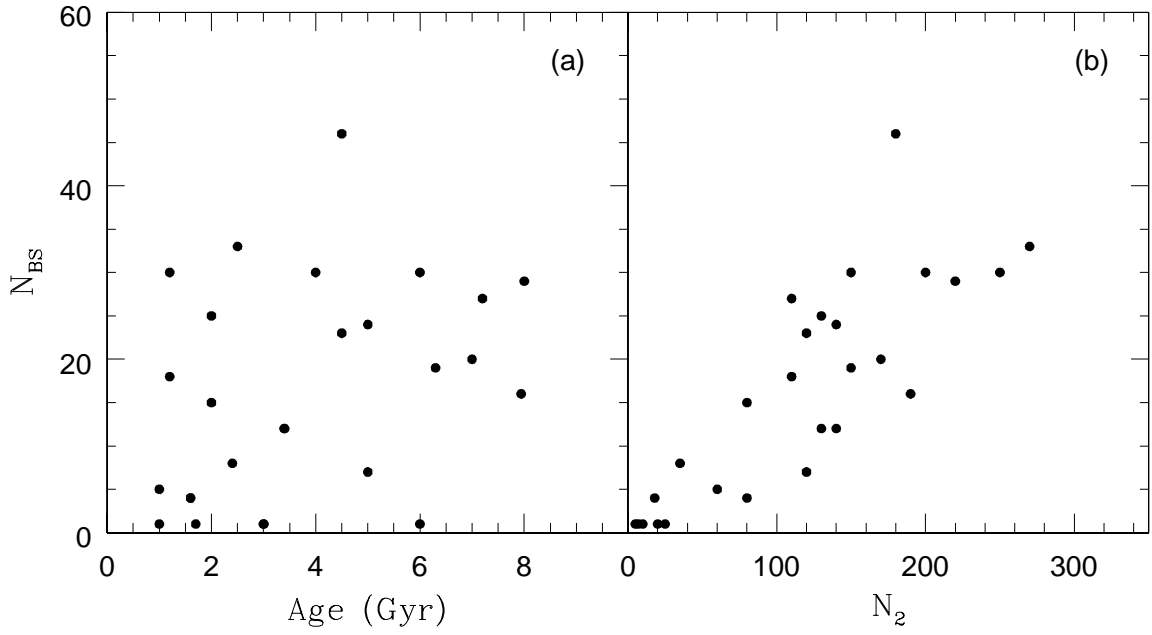}
\caption{(a): Number of BSSs per open cluster in cluster 
age interval. (b): Number of BSSs per open cluster in N$_2$ interval.
\label{fig2}}
\end{figure}

\begin{figure}
\plotone{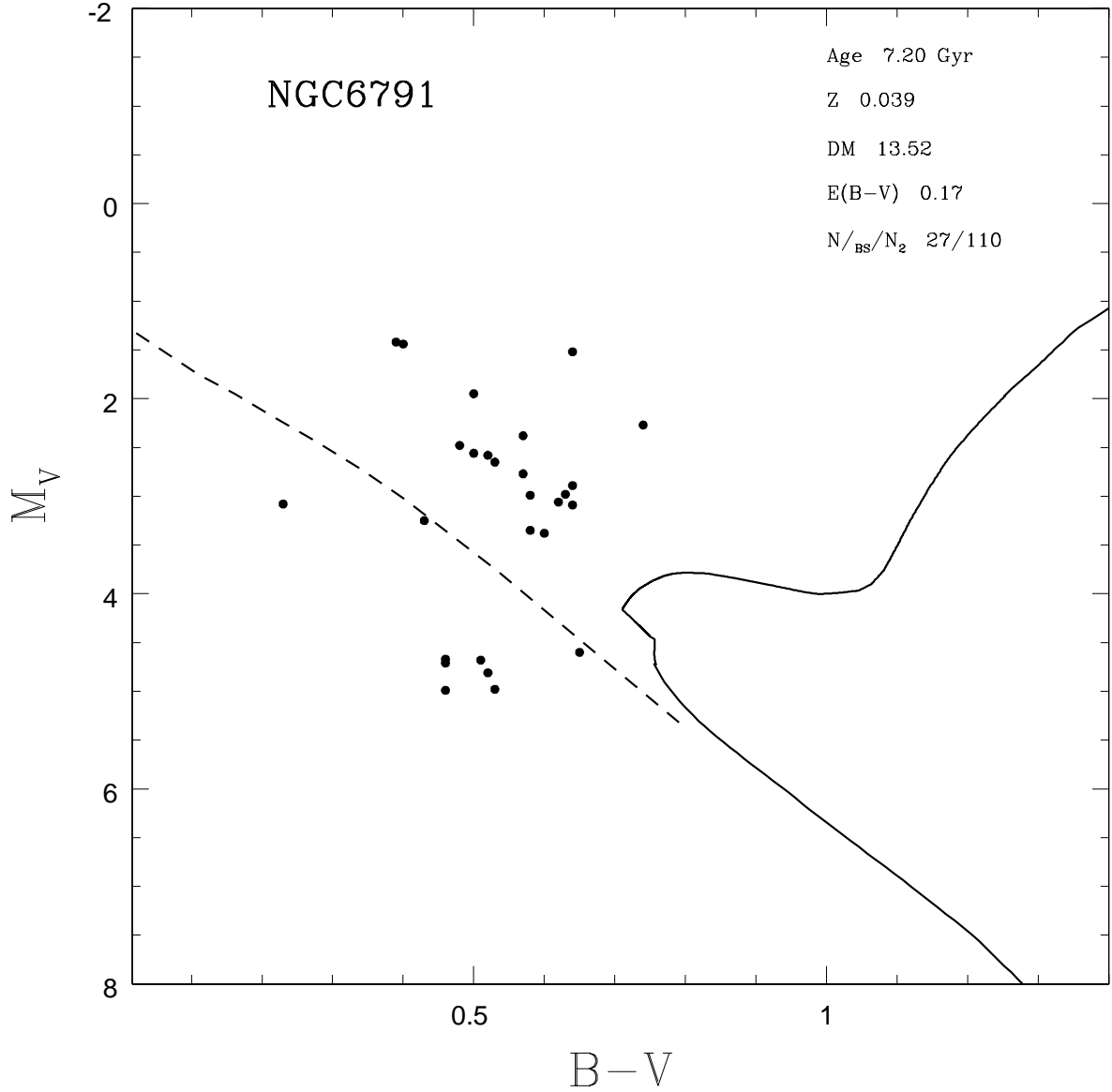}
\caption{The CMD of NGC 6791. $\it{Solid}$ $\it{line:}$
Theoretical isochrone according to the given age (7.2 Gyr) and 
metallicity (Z=0.039). $\it{Dash}$ $\it{line:}$ ZAMS. 
$\it{Solid}$ $\it{circles:}$ BSSs.\label{fig3}}
\end{figure}

\begin{figure}
\plotone{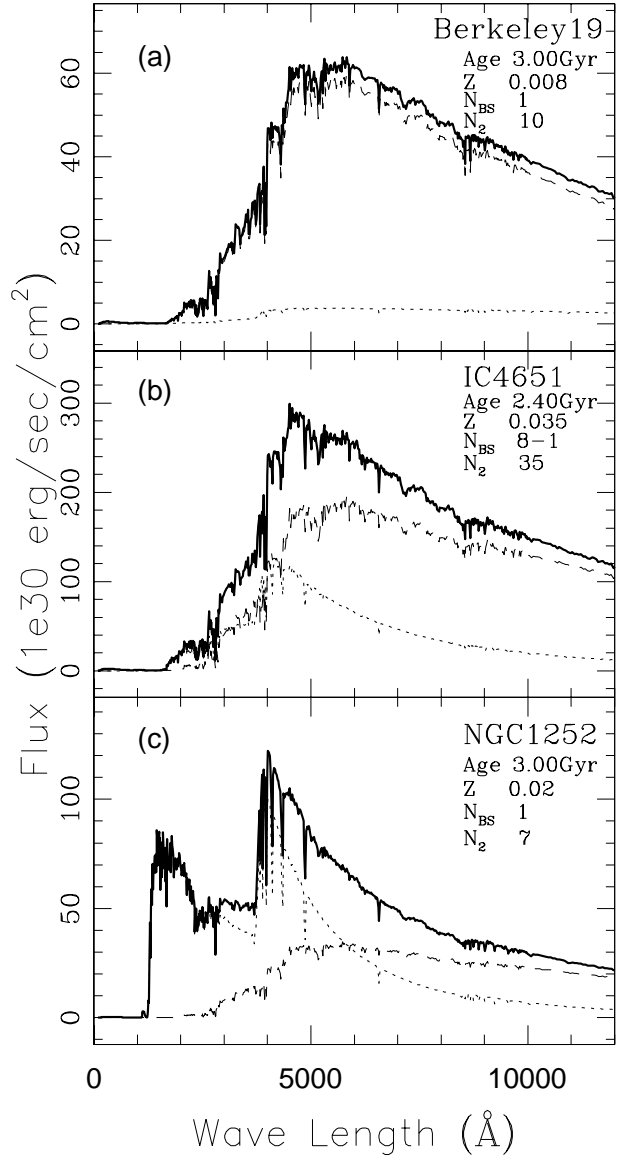}
\caption{The ISEDs modified by BSSs in different degrees. 
(a): Normal star dominate. (b): BSSs important. (c): BSSs dominate.
$\it{Dotted}$ $\it{line:}$ ISED of BSSs only. $\it{Dash}$ $\it{line:}$ 
ISED of isochrone only. $\it{Solid}$ $\it{line:}$ The composite ISED
of isochrone and BSSs.\label{fig4}}
\end{figure}

\begin{figure}
\plotone{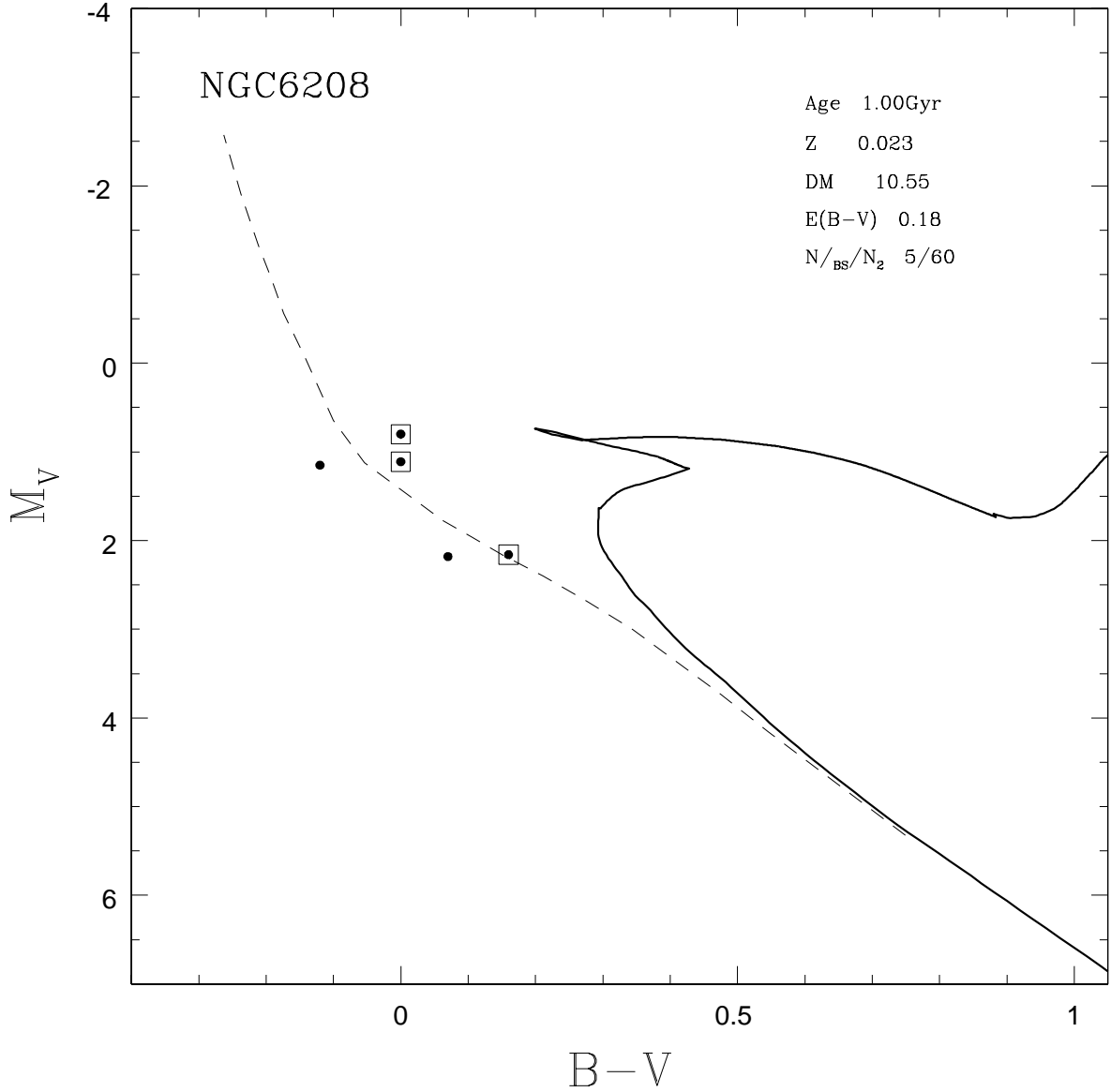}
\caption{The CMD of NGC 6208. $\it{Solid}$ $\it{line:}$ 
Theoretical isochrone.
$\it{Dash}$ $\it{line:}$ ZAMS. $\it{Solid}$ $\it{triangles:}$ Five BSSs of
NGC 6208 listed in AL95. $\it{Open}$ $\it{rectangles:}$ Three non-member
BSSs in Lindoff (1972).\label{fig5}}
\end{figure}

\begin{figure}
\plottwo{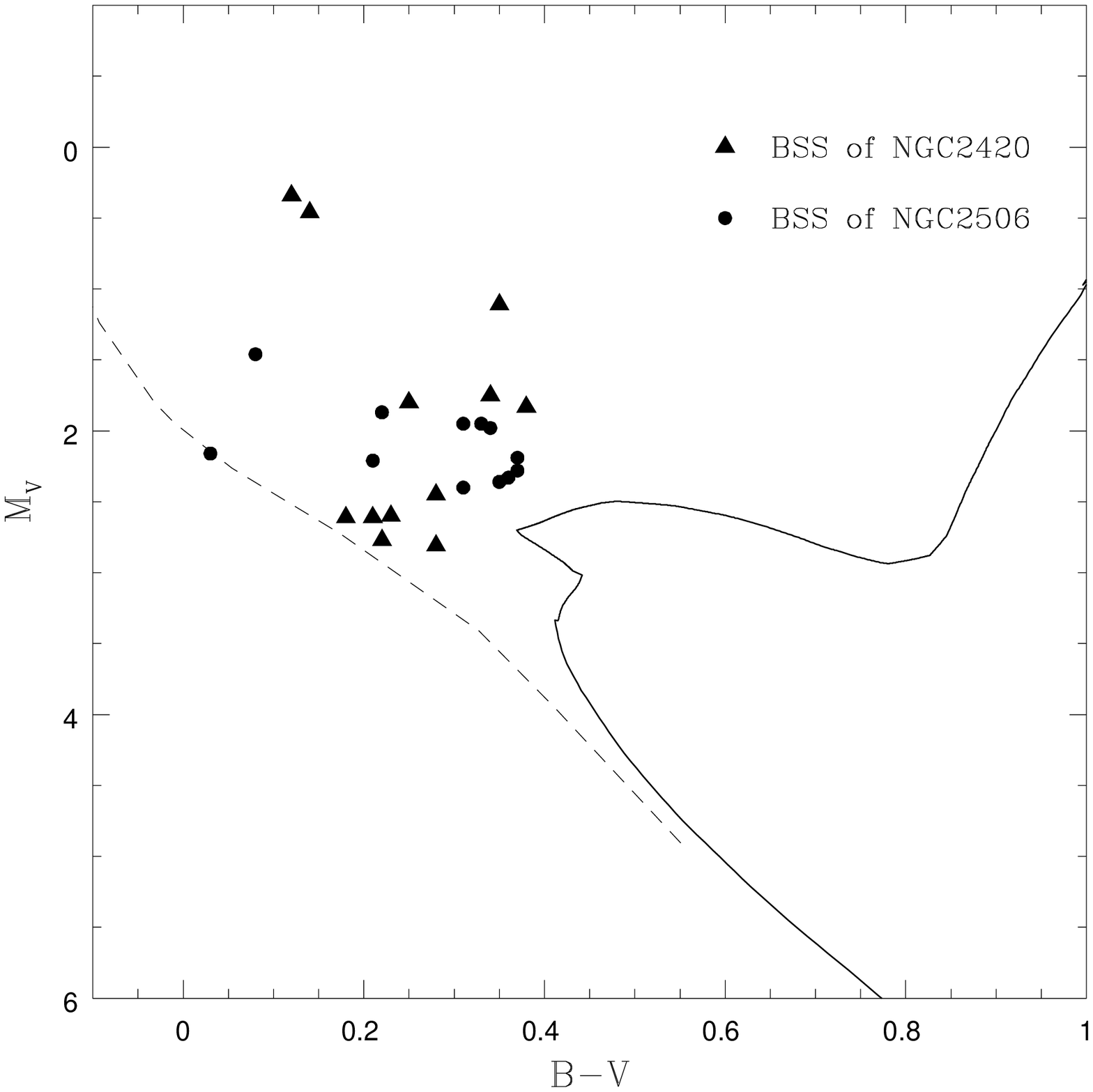}{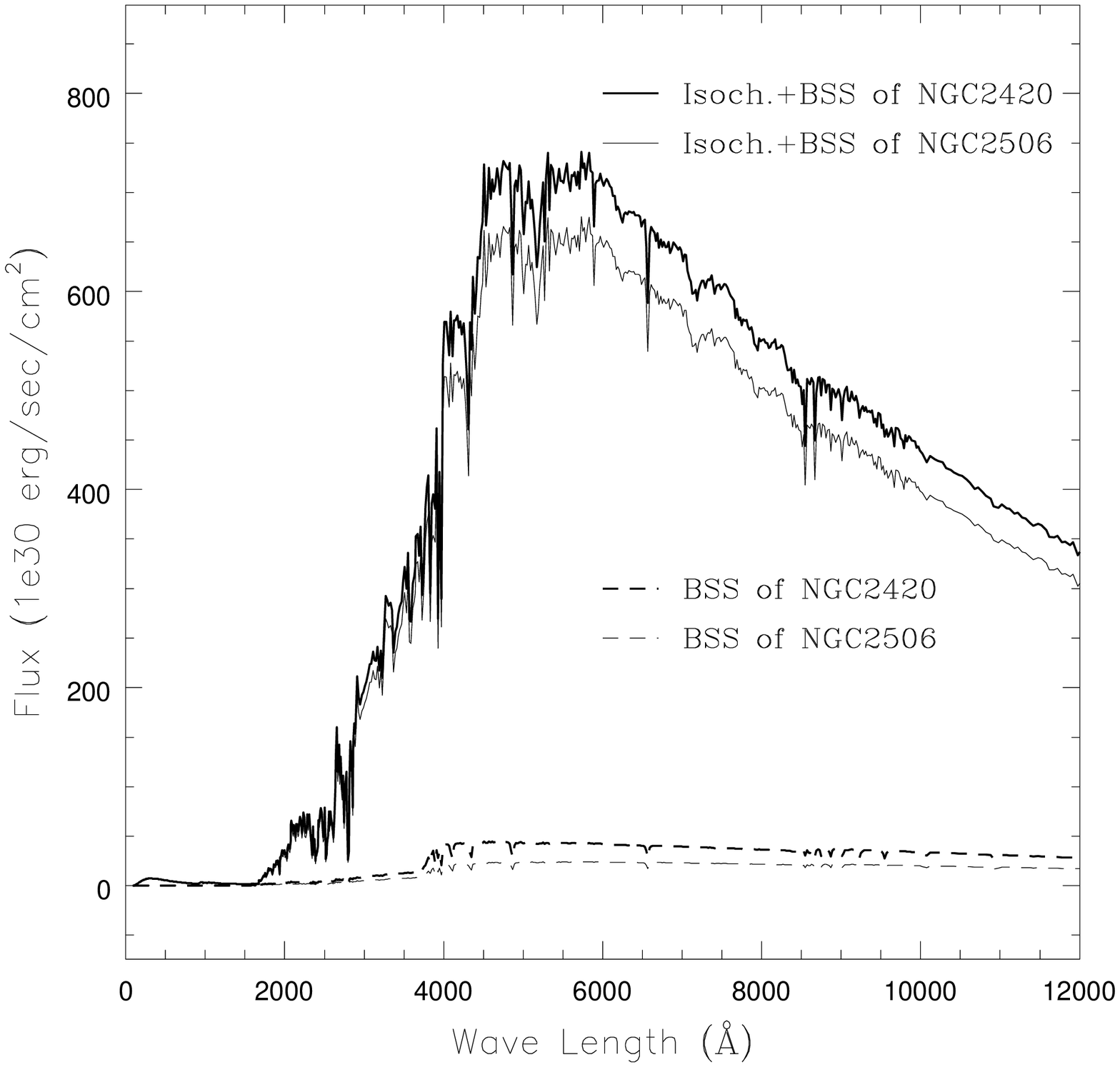}
\caption{A comparison of BSS distributions and 
ISED modifications
between NGC 2420 and NGC 2506. The left panel is the CMD, where
the $\it{Solid}$ $\it{line:}$
is the theoretical isochrone with age (3.4 Gyr) and metallicity (Z=0.007),
$\it{Dash}$ $\it{line:}$ is the ZAMS, $\it{Solid}$ $\it{circles:}$ 
are BSSs in NGC 2506, and $\it{Solid}$ $\it{triangles:}$ are BSSs in NGC 2420.
In the ISED plot (the right panel), 
$\it{Thick}$ $\it{solid}$ $\it{line:}$ is the ISED of isochrone plus BSSs in NGC 2420,
$\it{Thin}$ $\it{solid}$ $\it{line:}$ is the ISED of isochrone plus BSSs in NGC 2506,
$\it{Thick}$ $\it{dash}$ $\it{line:}$ is the ISED of BSSs in NGC 2420,
and $\it{Thin}$ $\it{dash}$ $\it{line:}$ is the ISED of BSSs in NGC 2506.\label{fig6}}
\end{figure}

\begin{figure}
\plotone{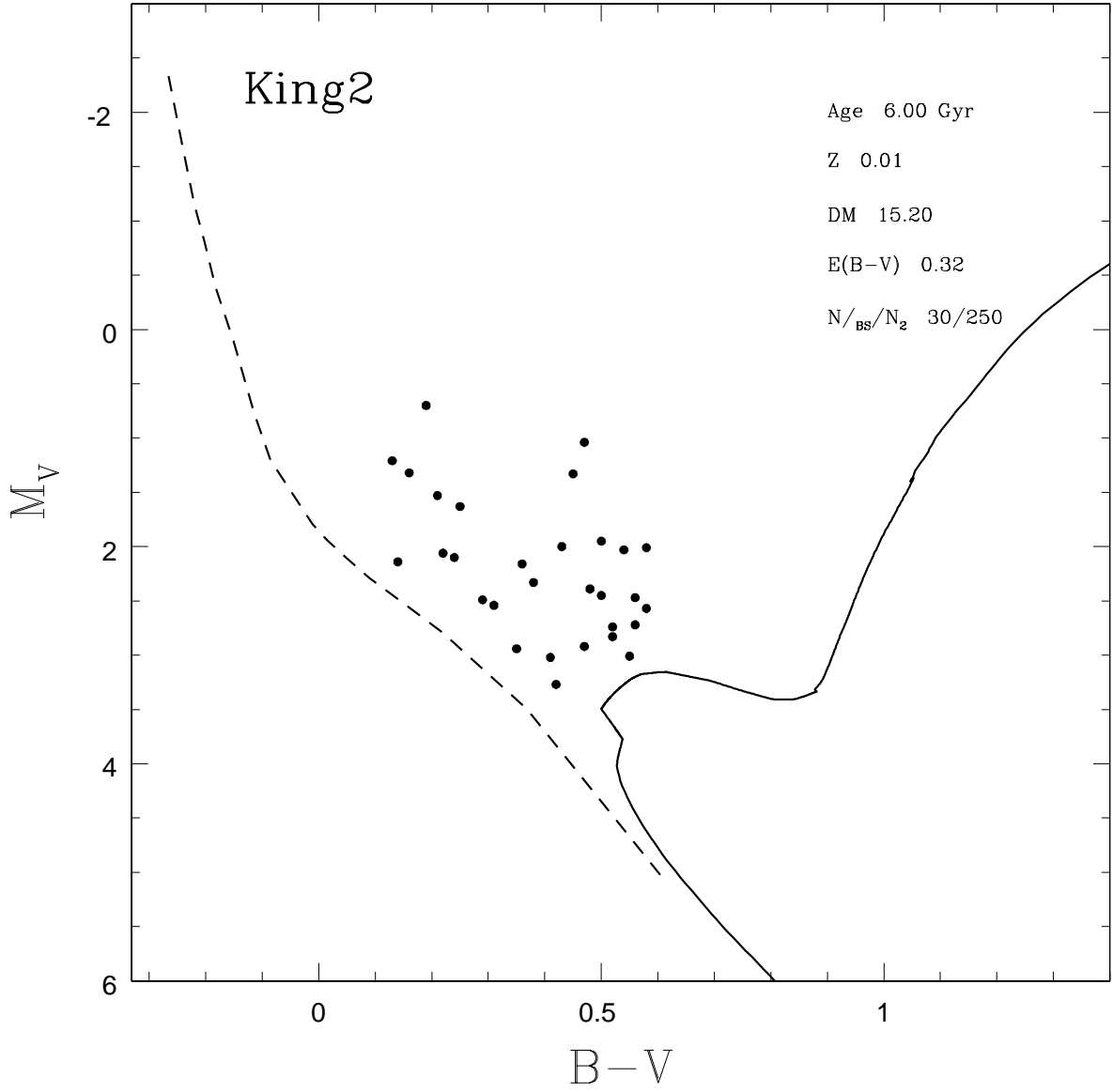}
\caption{The CMD of open cluster King 2.\label{fig7}}
\end{figure}

\begin{figure}
\plotone{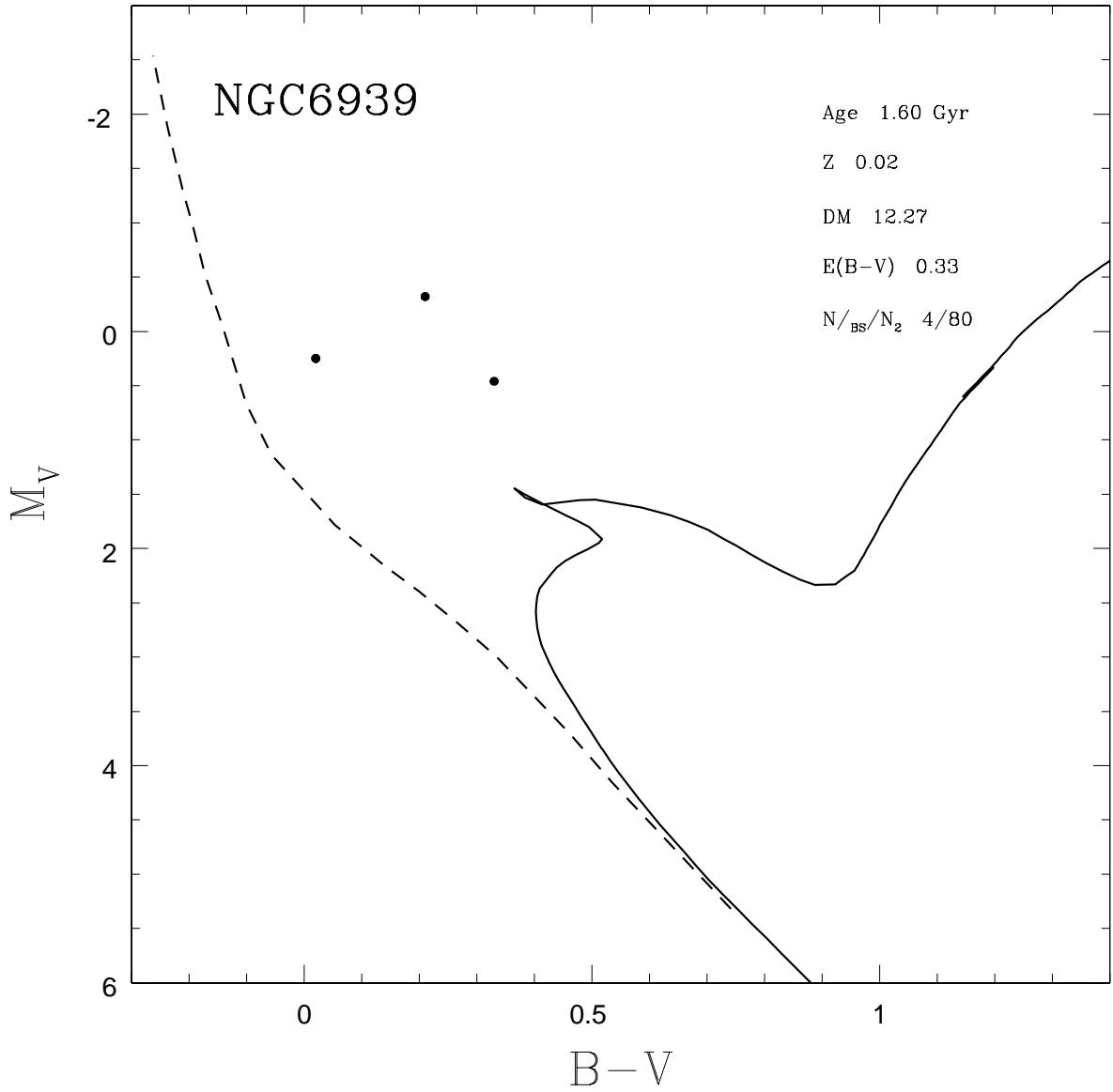}
\caption{The CMD of open cluster NGC 6939.\label{fig8}}
\end{figure}

\begin{figure}
\plottwo{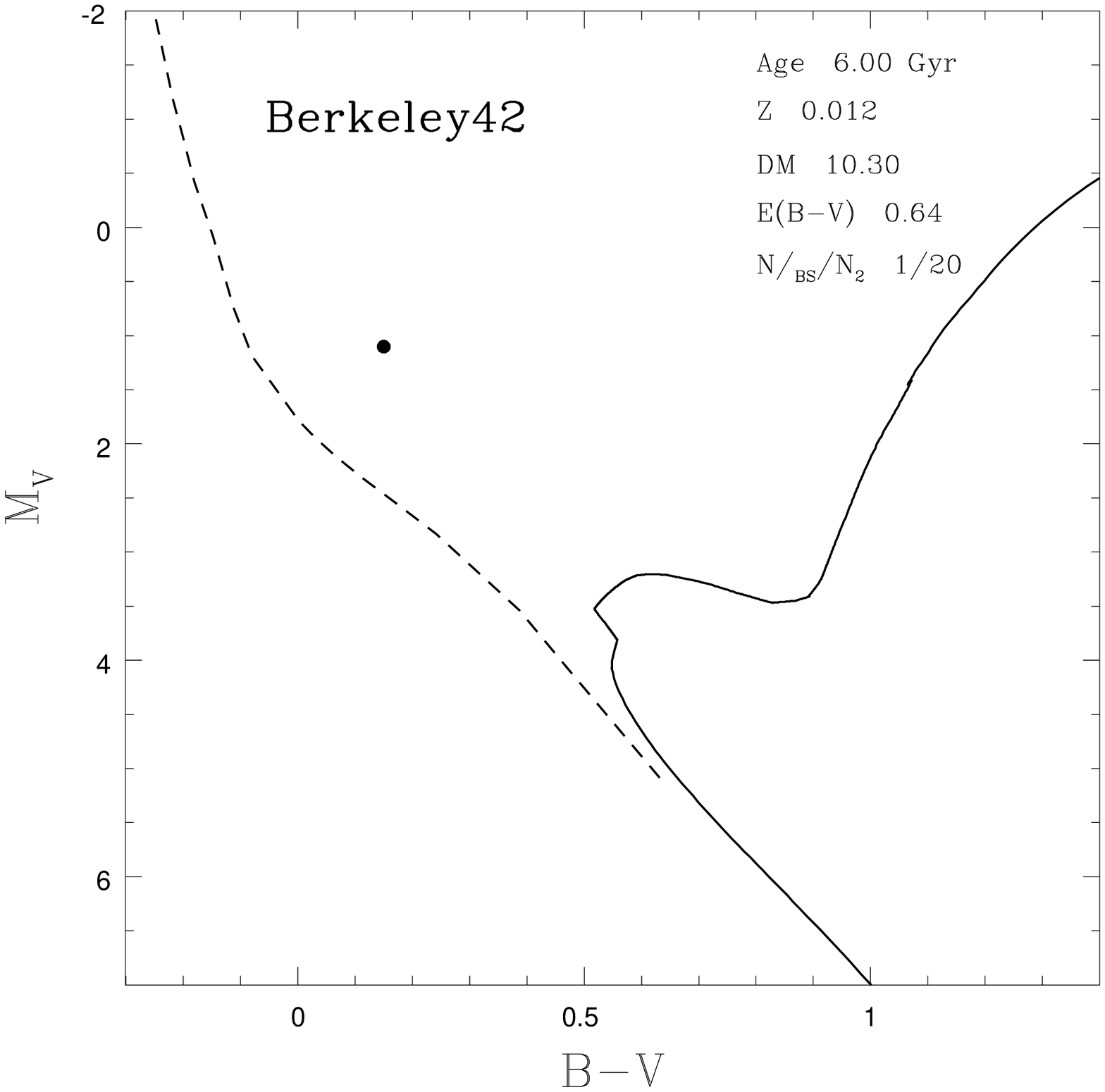}{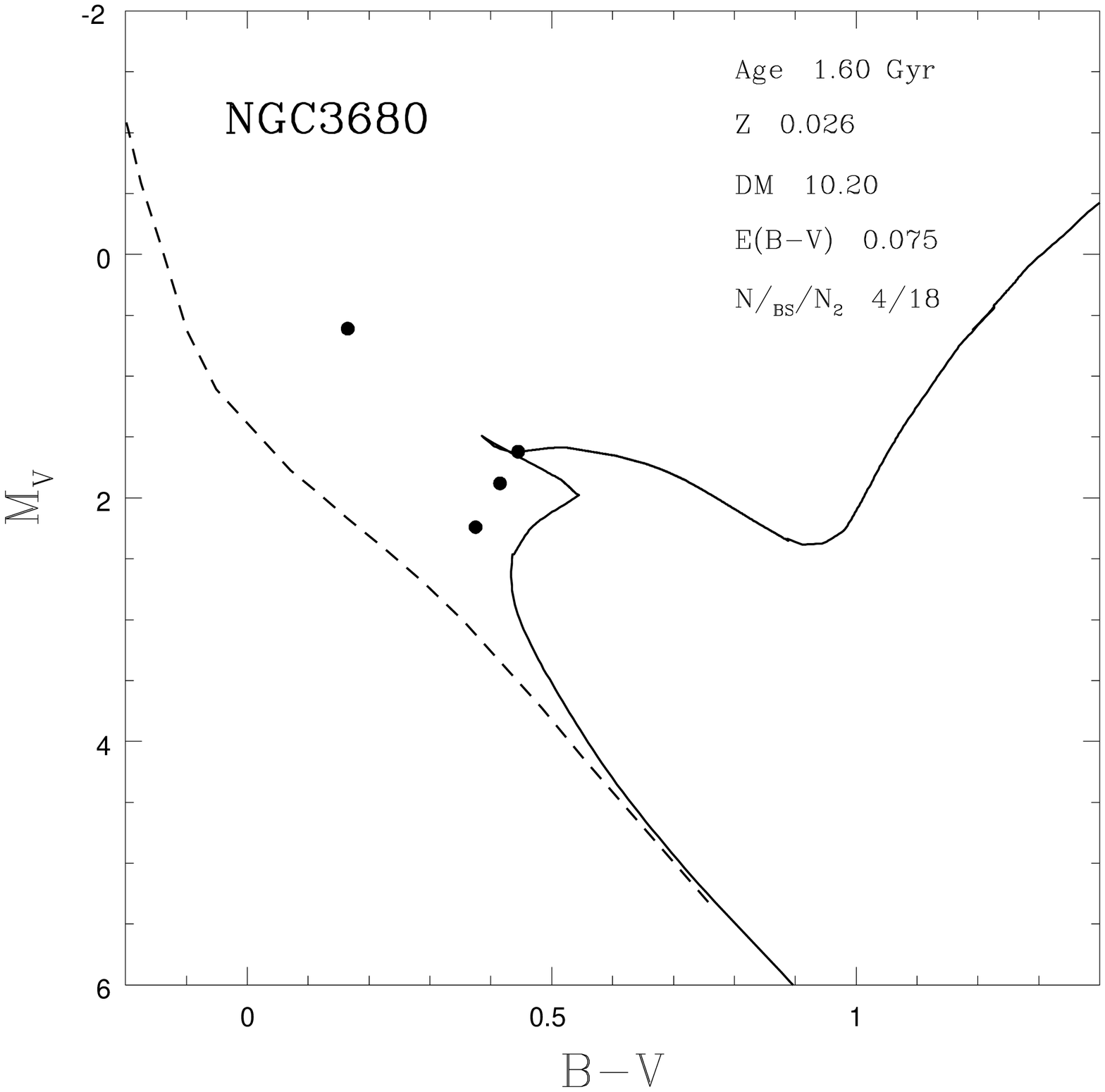}
\caption{CMDs of open cluster Berkeley 42 and NGC 3680.\label{fig9}}
\end{figure}

\begin{figure}
\plotone{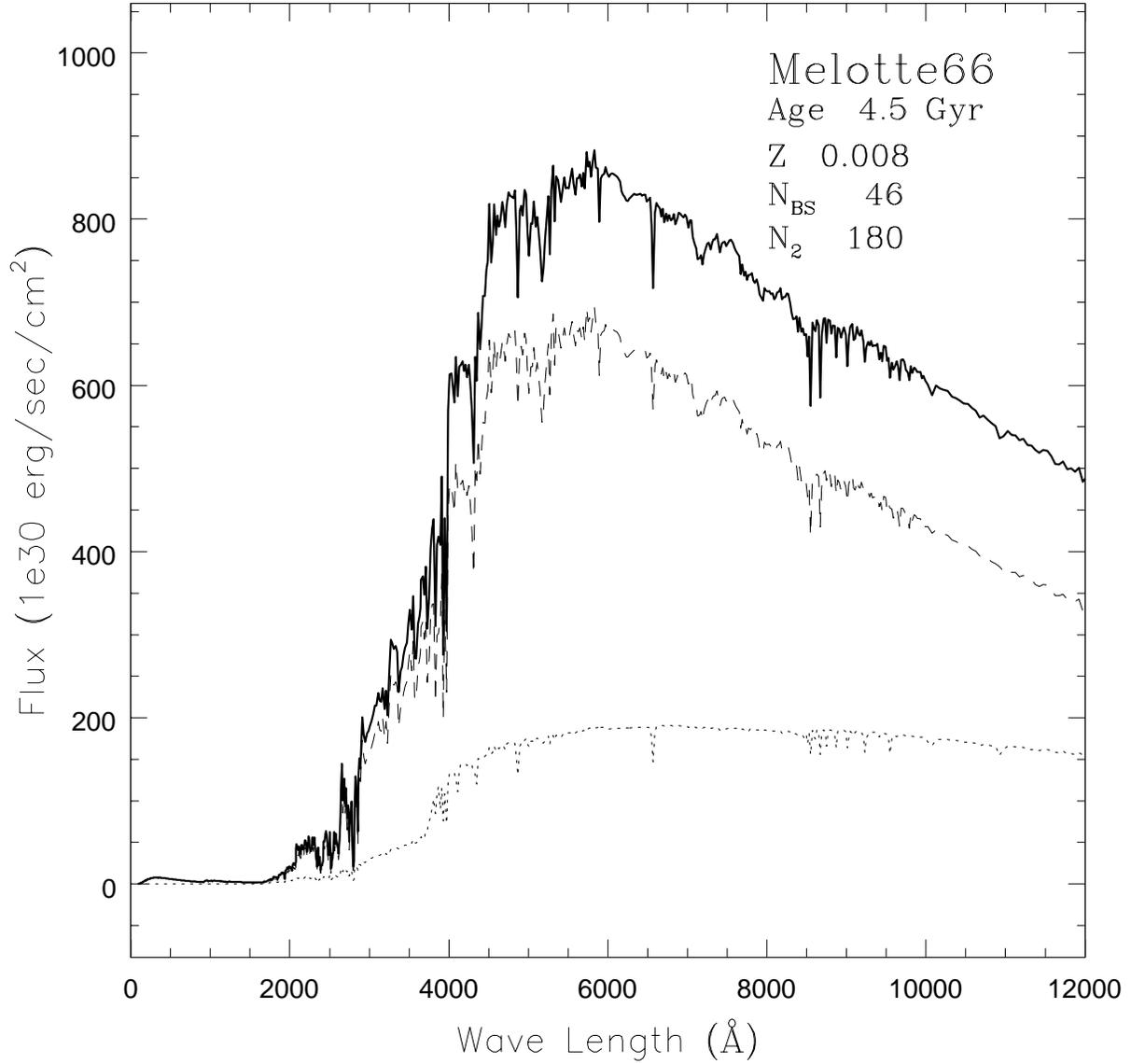}
\caption{The ISED of Melotte 66. $\it{Dotted}$ $\it{line:}$ ISED of BSSs only.
$\it{Dash}$ $\it{line:}$ ISED of isochrone only. $\it{Solid}$
$\it{line:}$ The composite ISED of isochrone and BSSs. 
\label{fig10}}
\end{figure}

\begin{figure}
\plotone{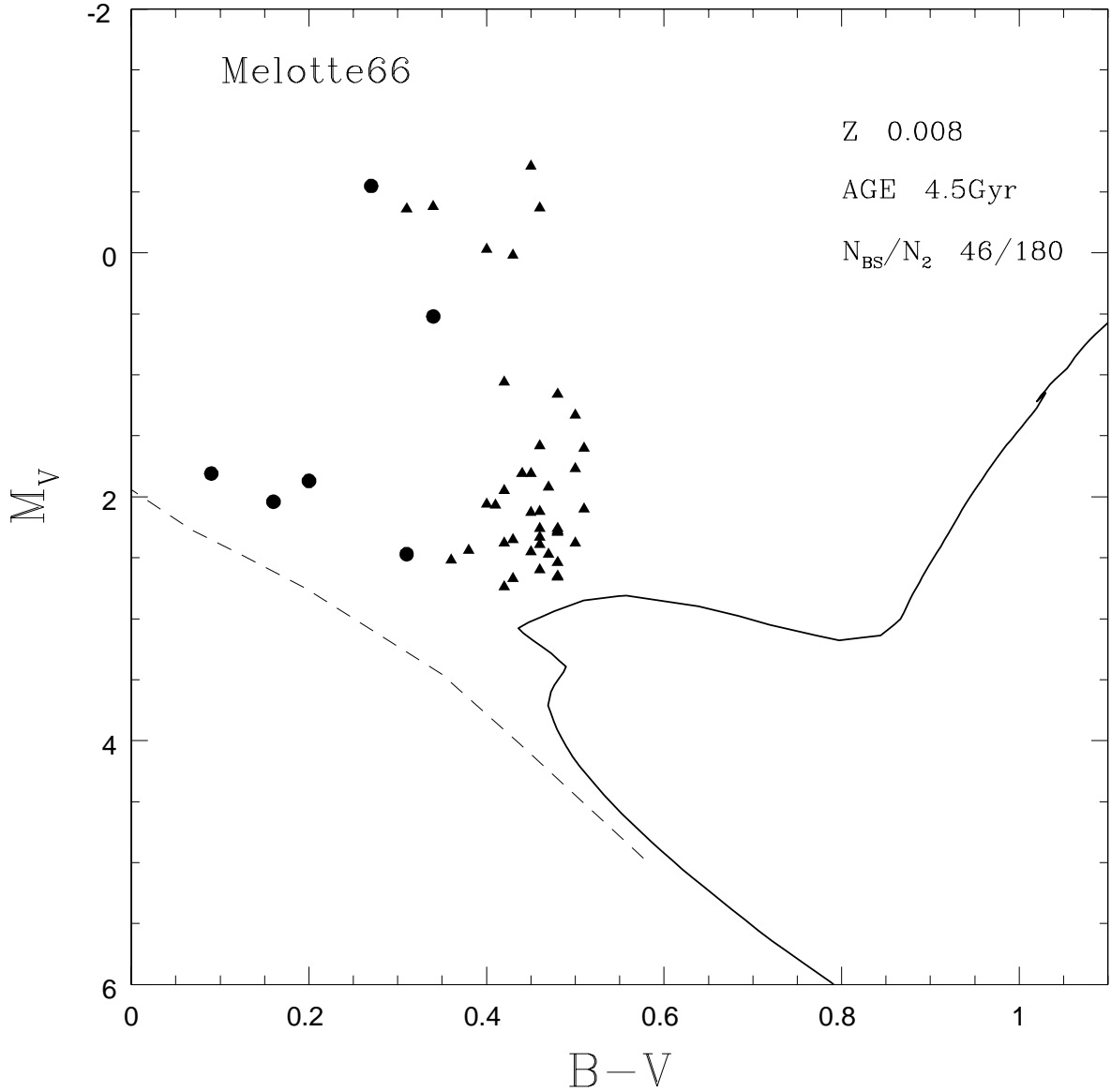}
\caption{The CMD of Melotte 66. $\it{Solid}$ $\it{line:}$
Theoretical isochrone according to the given age (8 Gyr) and metallicity
(Z=0.008). $\it{Dash}$ $\it{line:}$ ZAMS. $\it{Solid}$ $\it{circles:}$ 
Six blue-spectrum BSSs in the cluster. $\it{Solid}$ $\it{triangles:}$ 
Forty red-spectrum BSSs.\label{fig11}}
\end{figure}

\begin{figure}
\plottwo{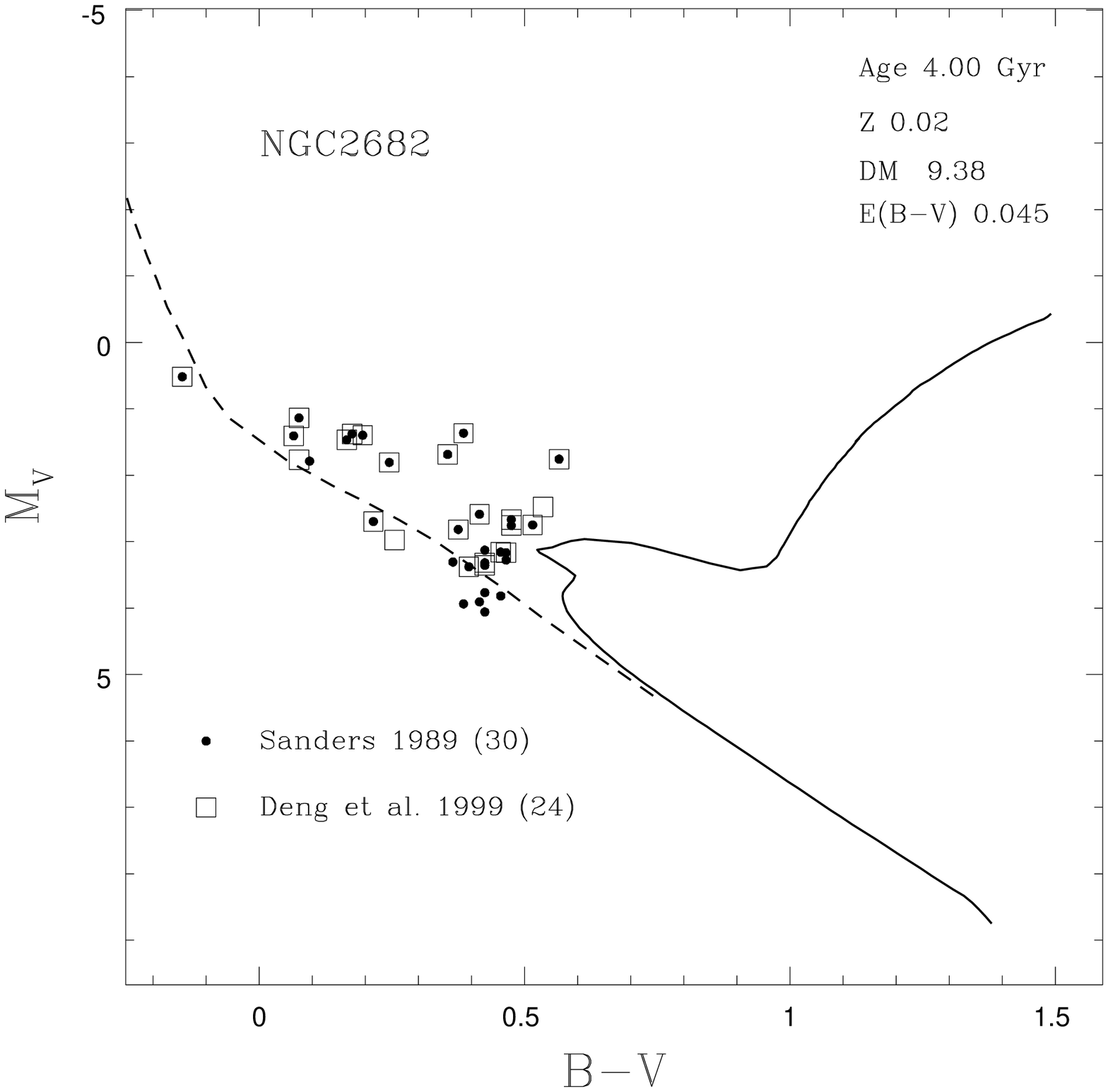}{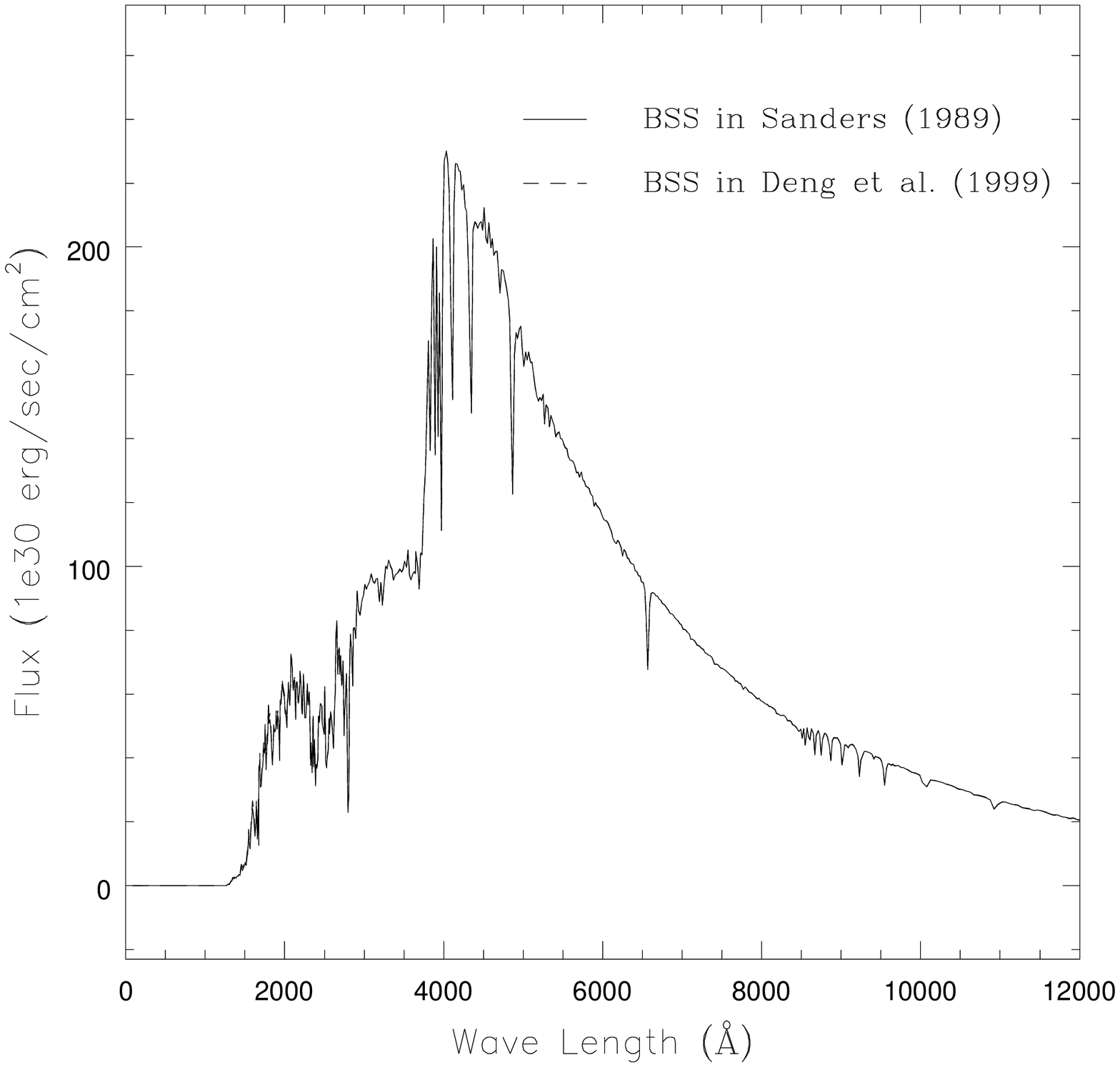}
\caption{Comparison with BSSs identifications in two work of NGC 2682. 
In the CMD, $\it{solid}$ $\it{dots}$ are BSSs in Sanders (1989),
and $\it{open}$ $\it{rectangles}$ are BSSs in Deng et al. (1999).
In the ISED, $\it{solid}$ $\it{line}$ is the integrated light of BSSs 
in Sanders (1989), and $\it{dash}$ $\it{line}$ is that in Deng et al. (1999).
\label{fig12}}
\end{figure}

\clearpage

\begin{figure}
\plottwo{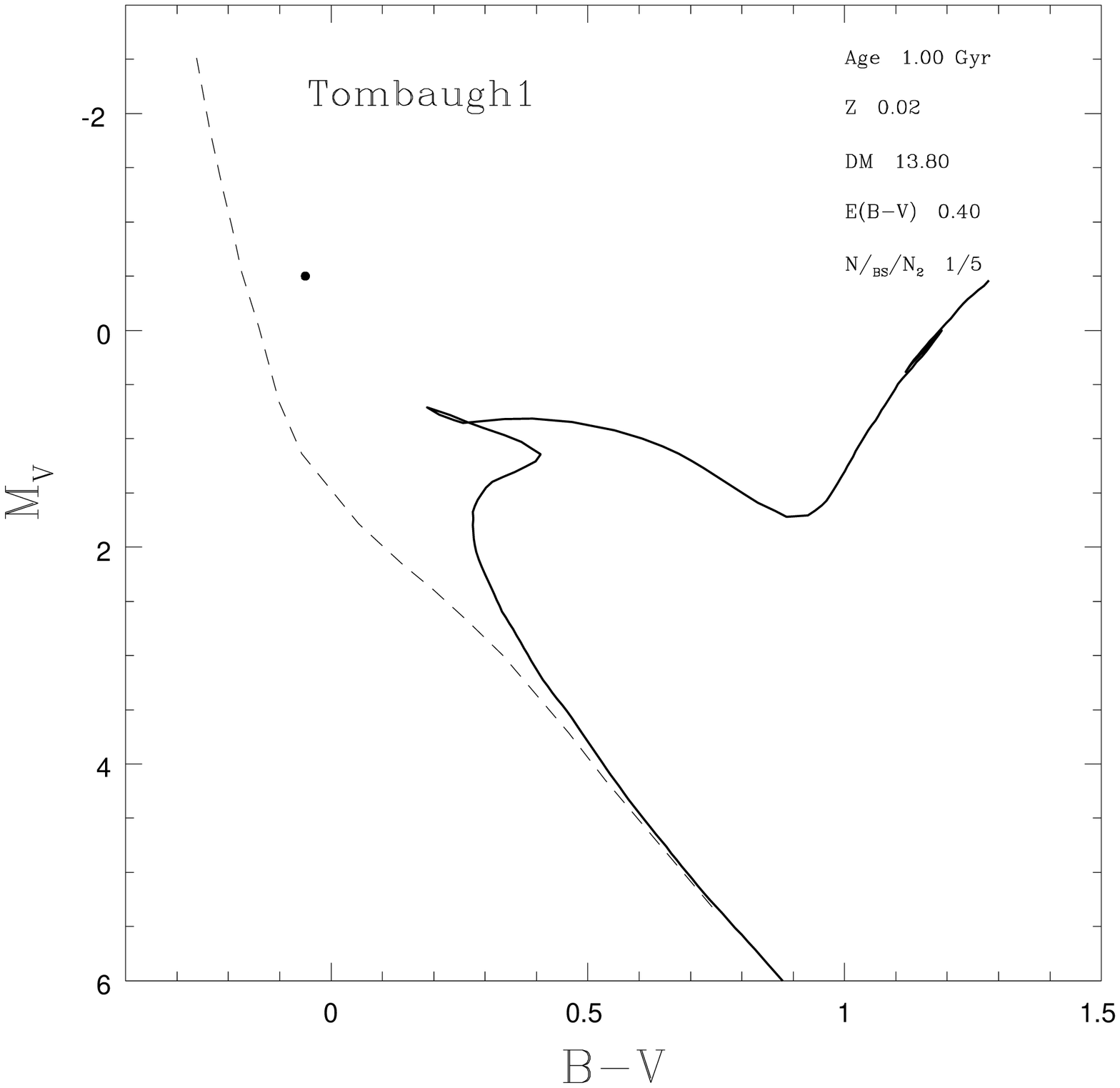}{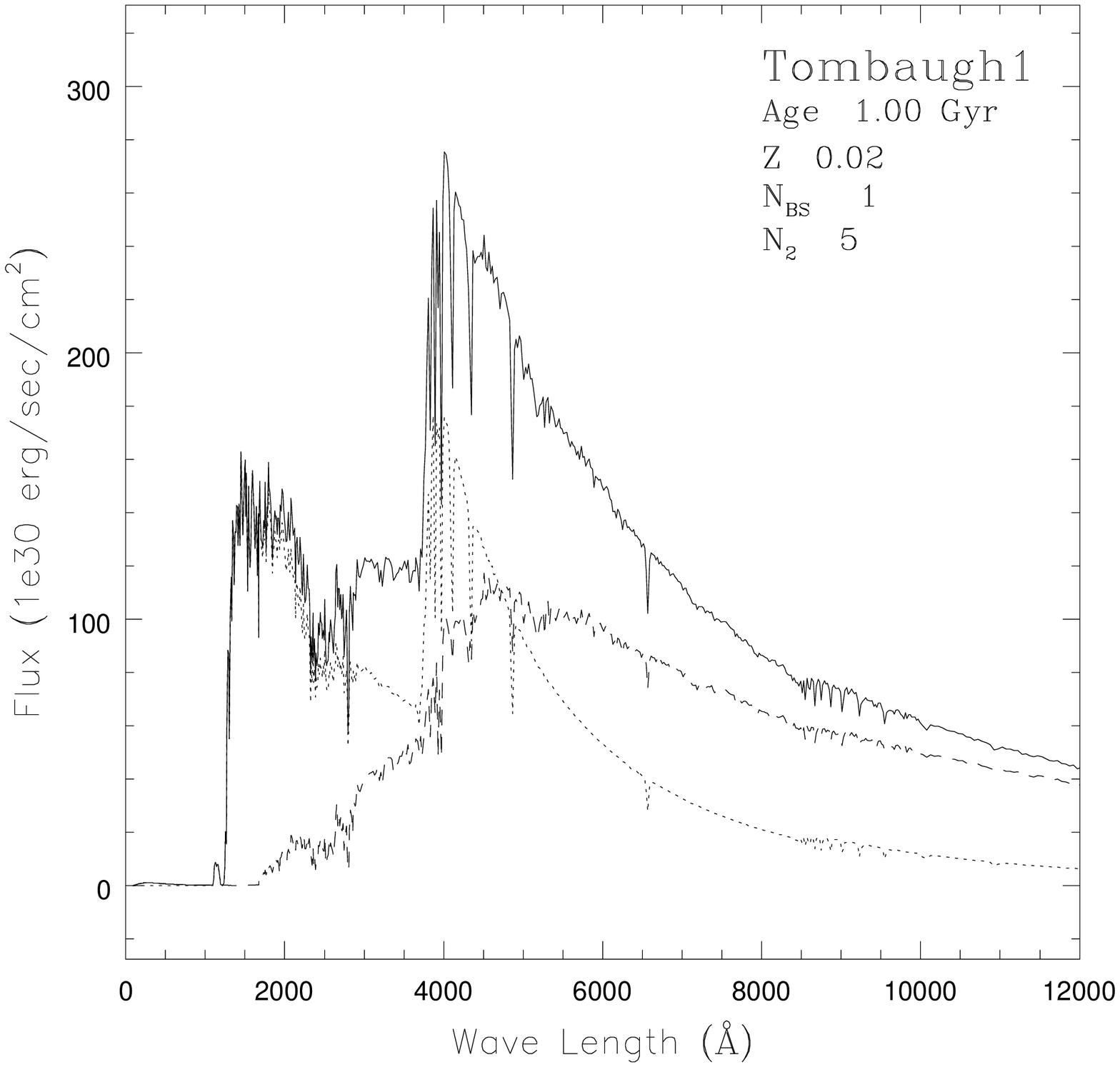}
\caption{The CMD and ISED of Tombaugh 1.\label{fig13}}
\end{figure}

\begin{figure}
\plotone{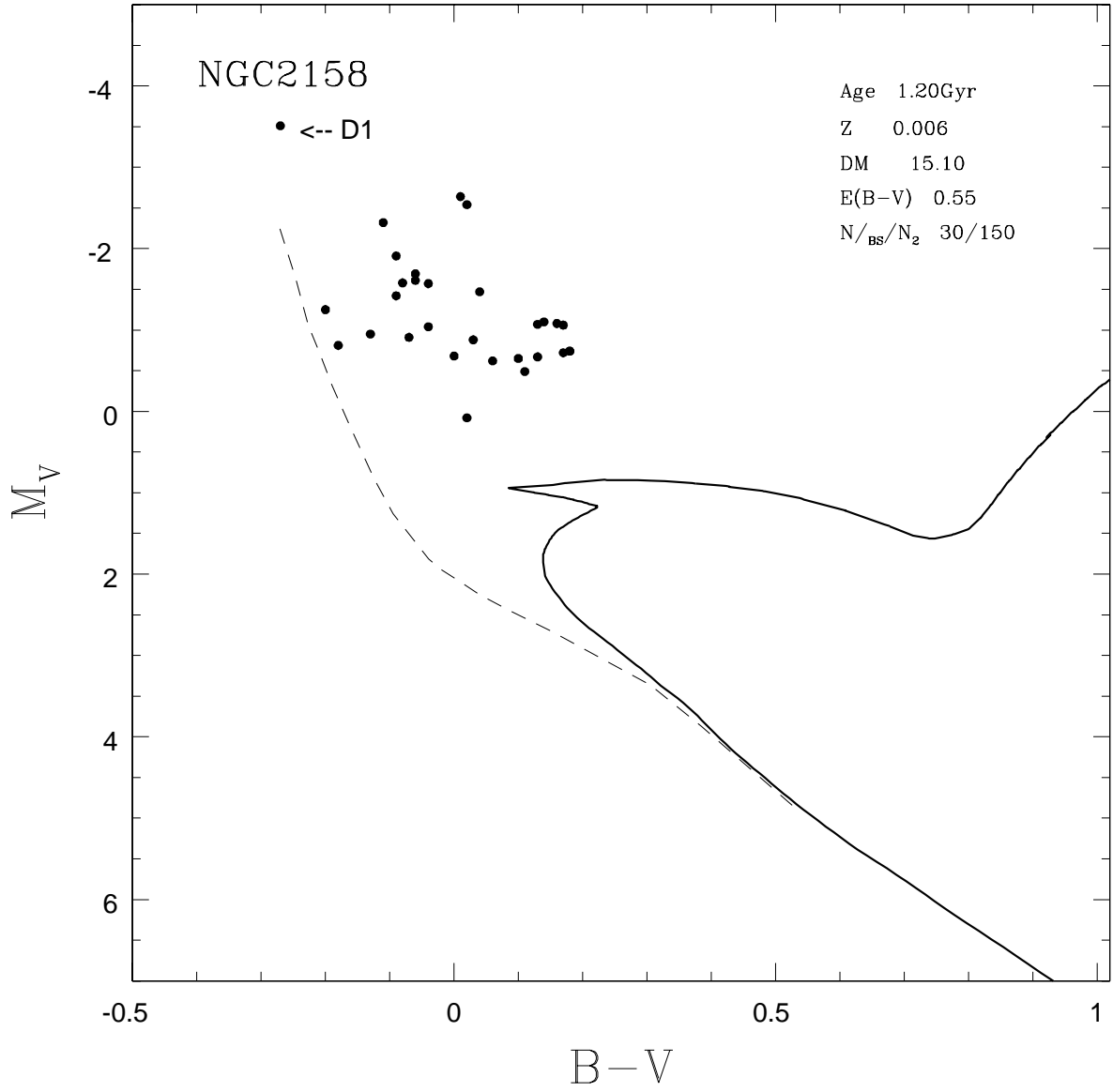}
\caption{The CMD of NGC 2158. $\it{Solid}$ $\it{line:}$
Theoretical isochrone with age (1.2 Gyr) and metallicity (Z=0.008).
$\it{Dash}$ $\it{line:}$ ZAMS. $\it{Solid}$ $\it{circles:}$ BSSs.
'D1' is the brightest BSS in the cluster.\label{fig14}} 
\end{figure}

\begin{figure}
\plotone{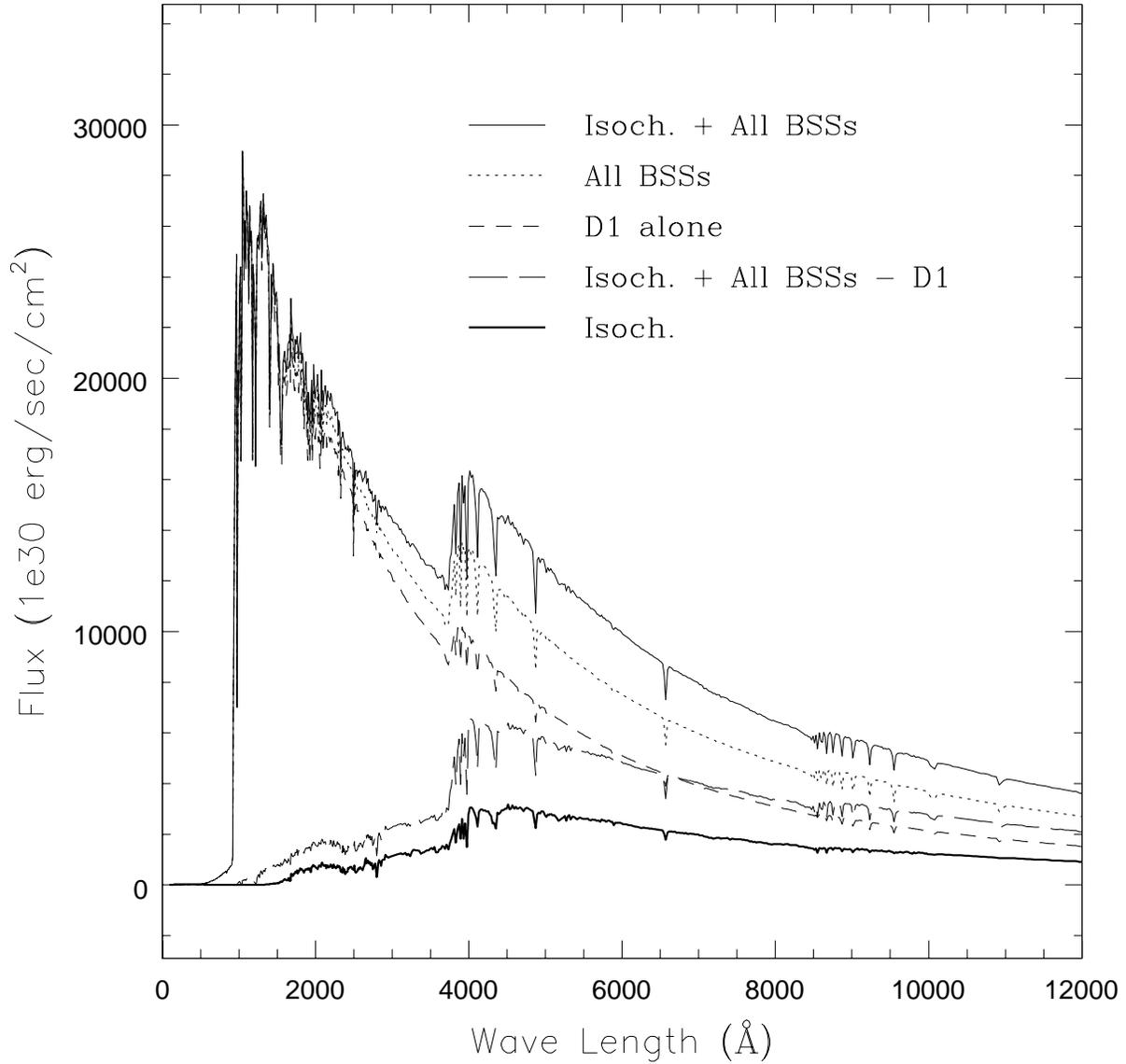}
\caption{Integrated flux of NGC 2158. The contribution of BSSs population is
taken apart. $\it{Thick}$ $\it{solid}$ $\it{line:}$ ISED of isochrone. 
$\it{Long}$ $\it{dash}$ $\it{line:}$ All components except D1. 
$\it{Short}$ $\it{dash}$ $\it{line:}$ D1 alone. 
$\it{Thin}$ $\it{solid}$ $\it{line:}$ All components (Isochrone and all BSSs).
\label{fig15}}
\end{figure}

\begin{figure}
\plotone{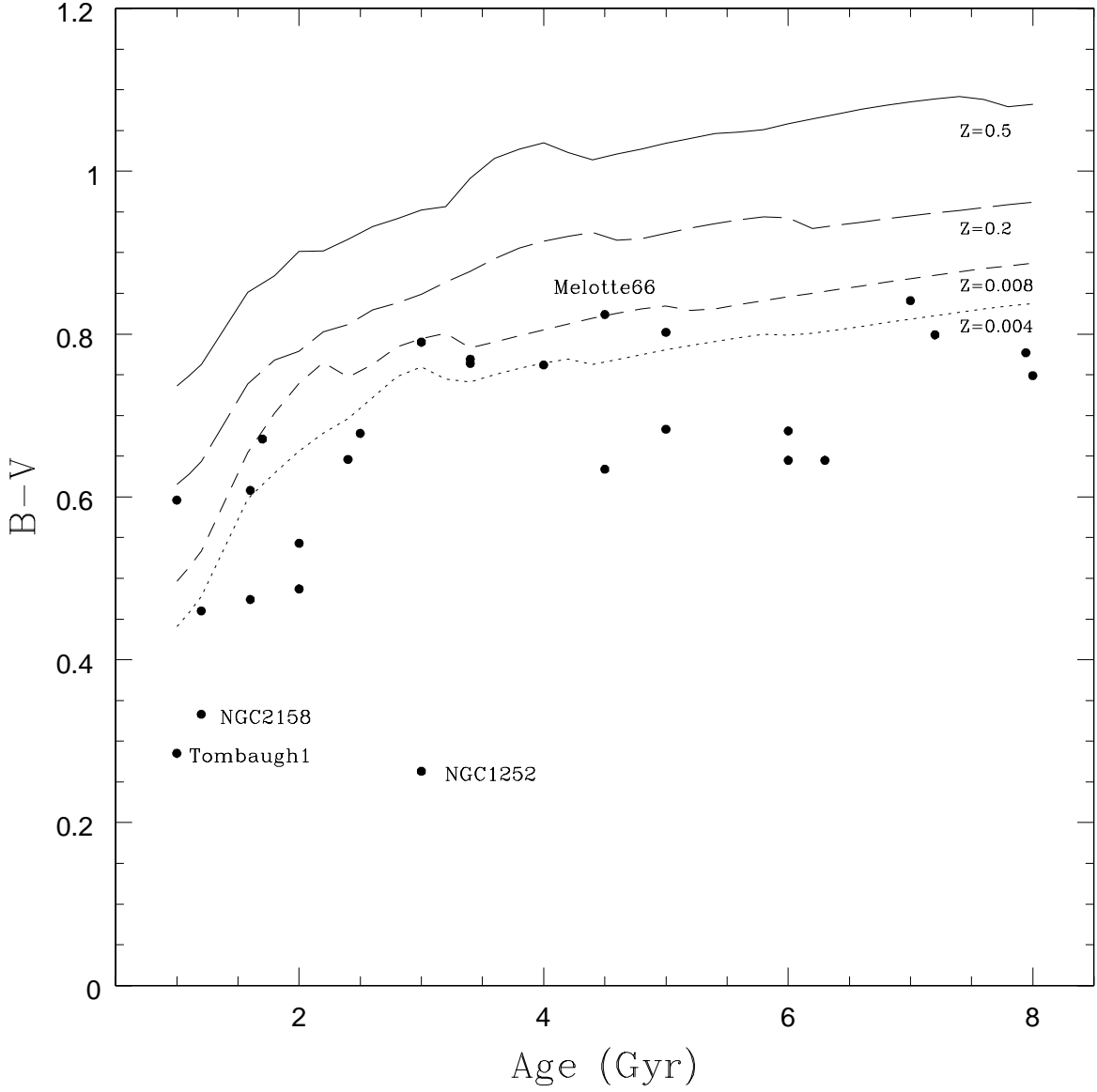}
\caption{B-V color modified by BSSs. Four lines are theoretical values 
according to different metallicities. The $\it{Solid}$ $\it{circles}$
are B-V color of our sample clusters involving BSSs contribution.
\label{fig16}} 
\end{figure}

\begin{figure}
\plotone{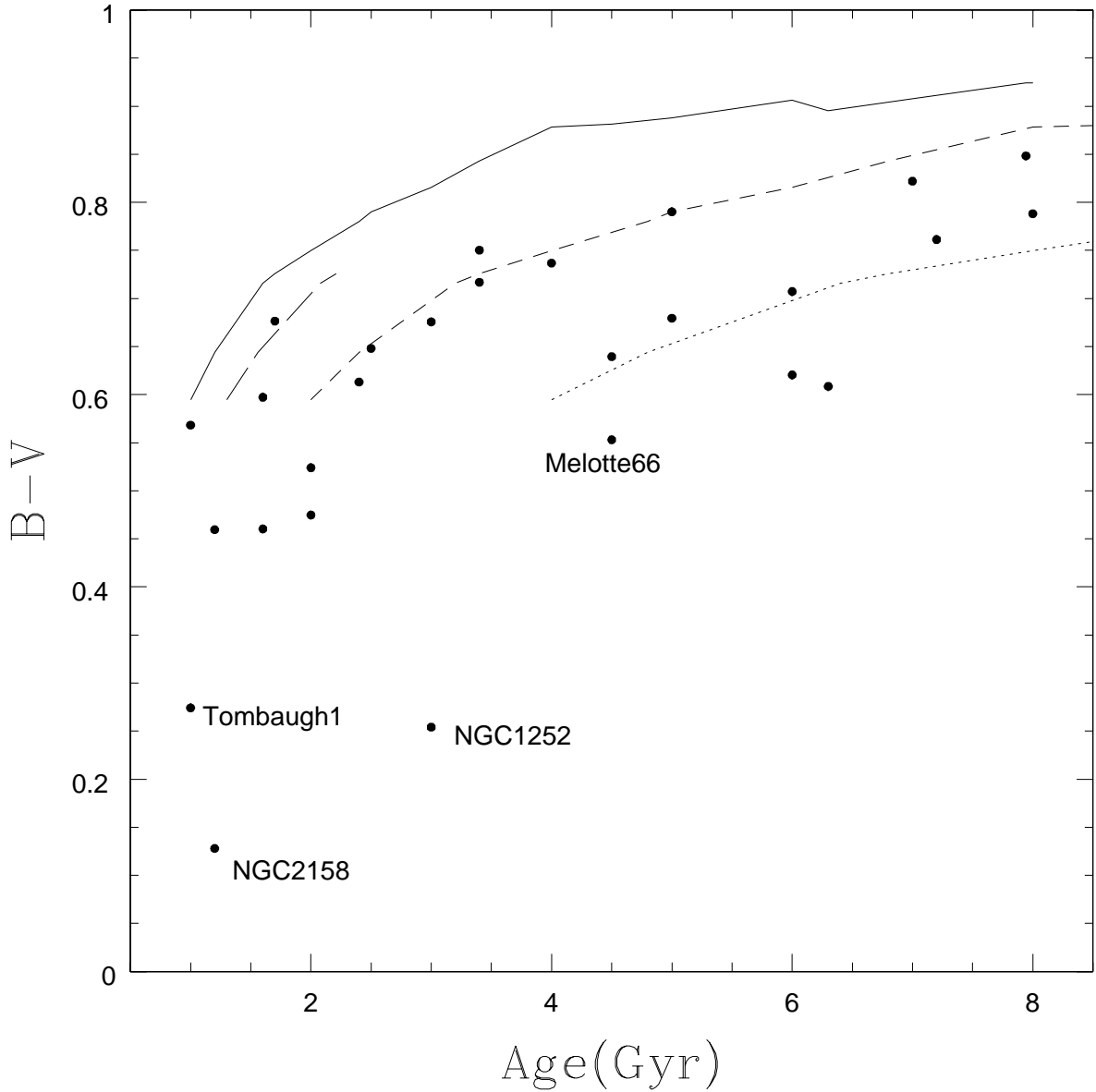}
\caption{Detecting age underestimation of star clusters when fitting 
the observed color with conventional SSP model. 
Assuming all the sample clusters with solar metallicity, 
the $\it{solid}$ $\it{line}$ is theoretical B-V color. Keeping B-V values
and double age is the $\it{Dash}$ $\it{line}$, and triple age is the
$\it{dotted}$ $\it{line}$. The $\it{long}$ $\it{dash}$ ${line}$ is 
plotted with age 30\%-expanding of the conventional value.\label{fig17}}
\end{figure}

\begin{figure}
\plotone{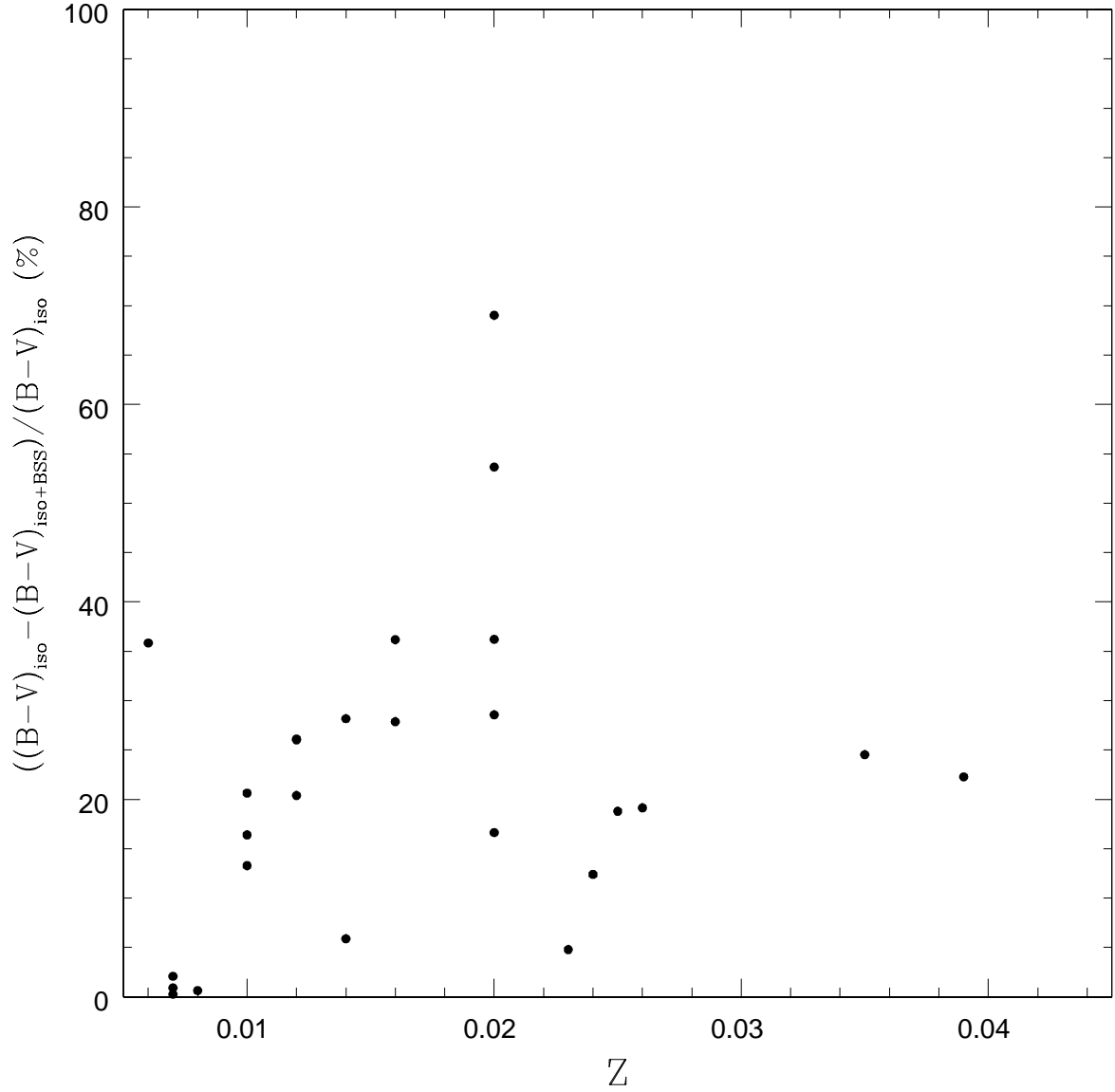}
\caption{Correlation between metallicity and the ratio of the
alternation of B-V color caused by BSSs to the conventional B-V
color.\label{fig18}}
\end{figure}

\begin{figure}
\plotone{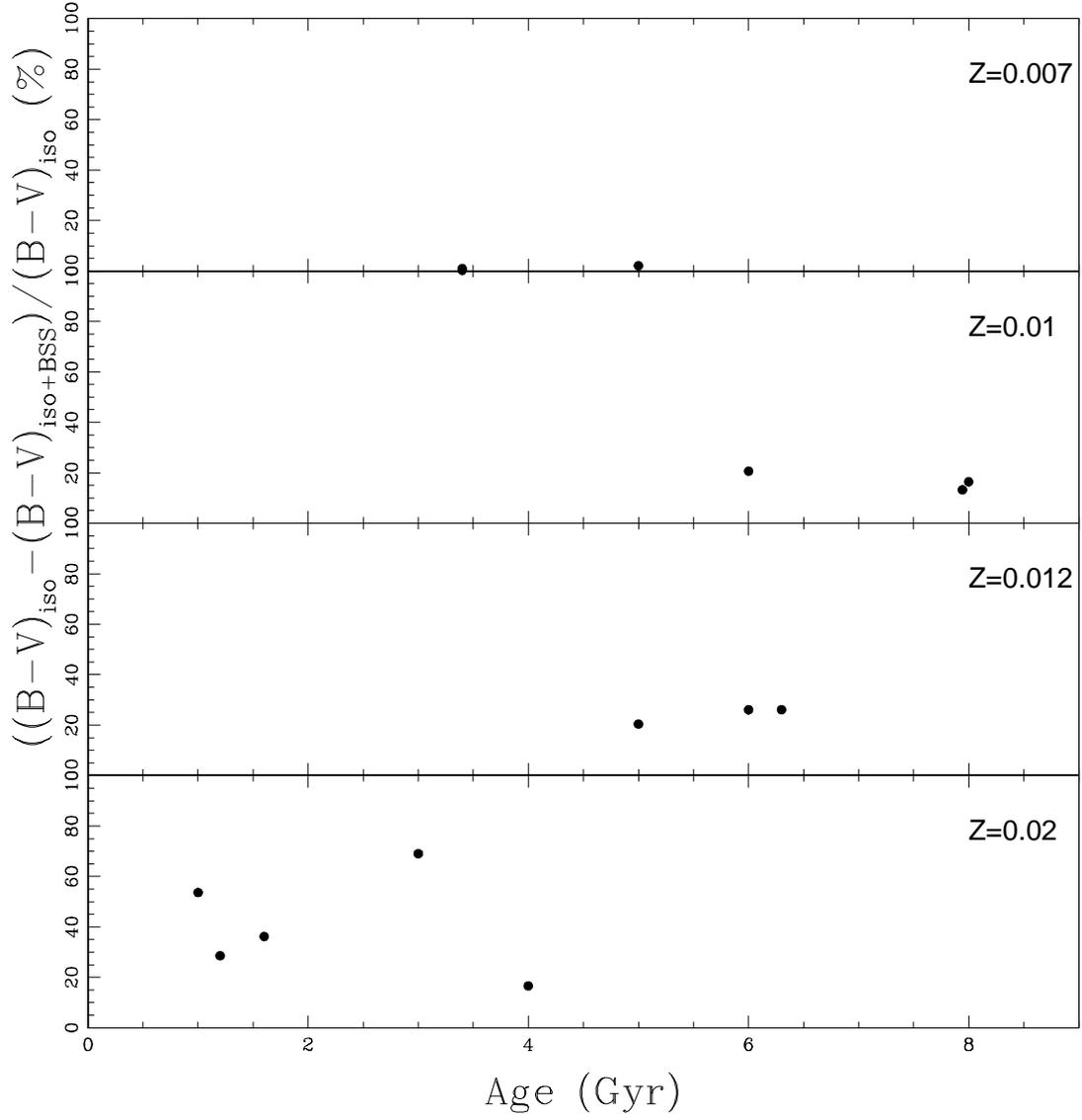}
\caption{BSSs contribution to cluster B-V color, presented with
four metallicities: Z=0.007, Z=0.01, Z=0.012 and Z=0.02.\label{fig19}}
\end{figure}

\clearpage

\begin{figure}
\plottwo{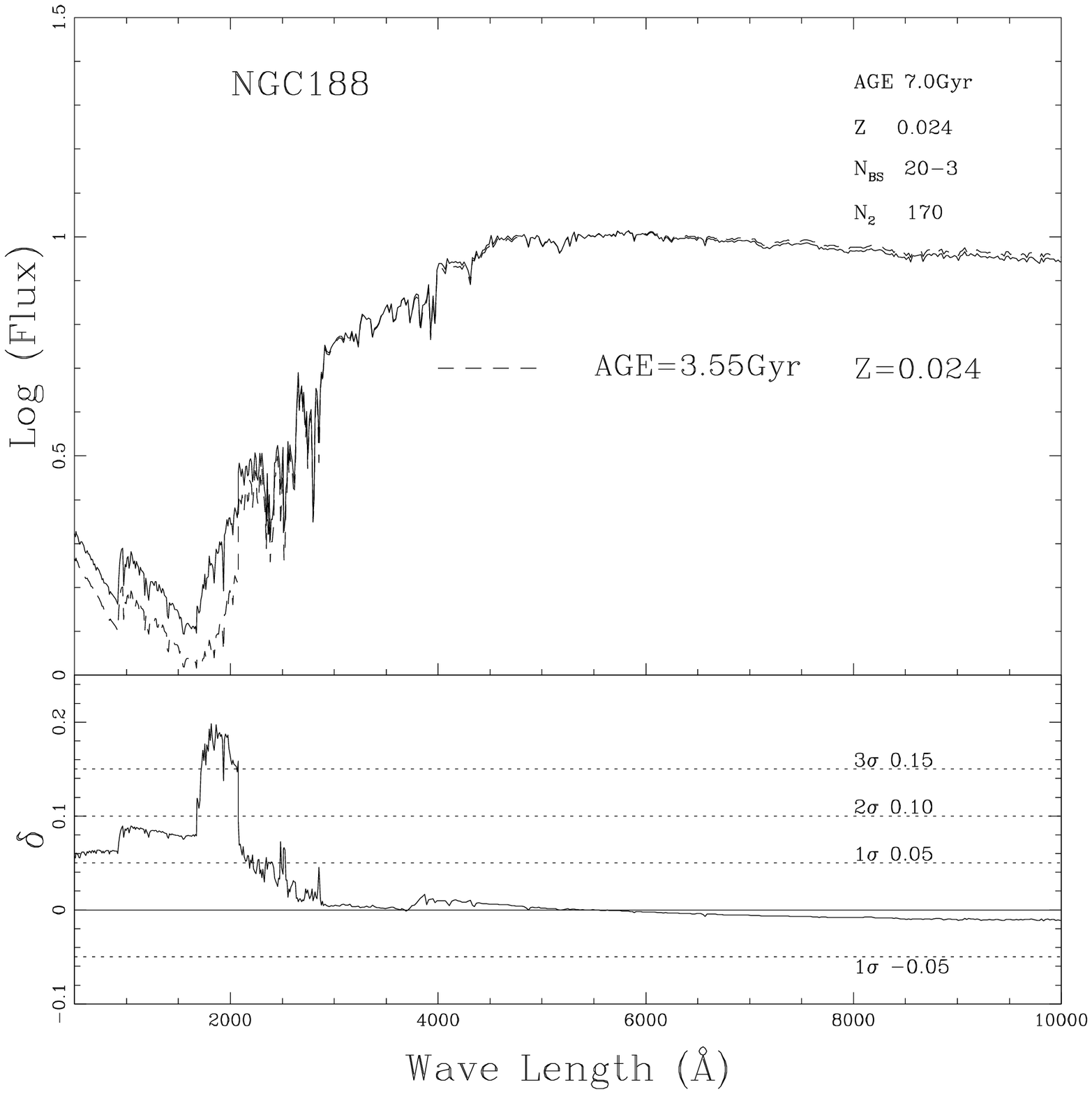}{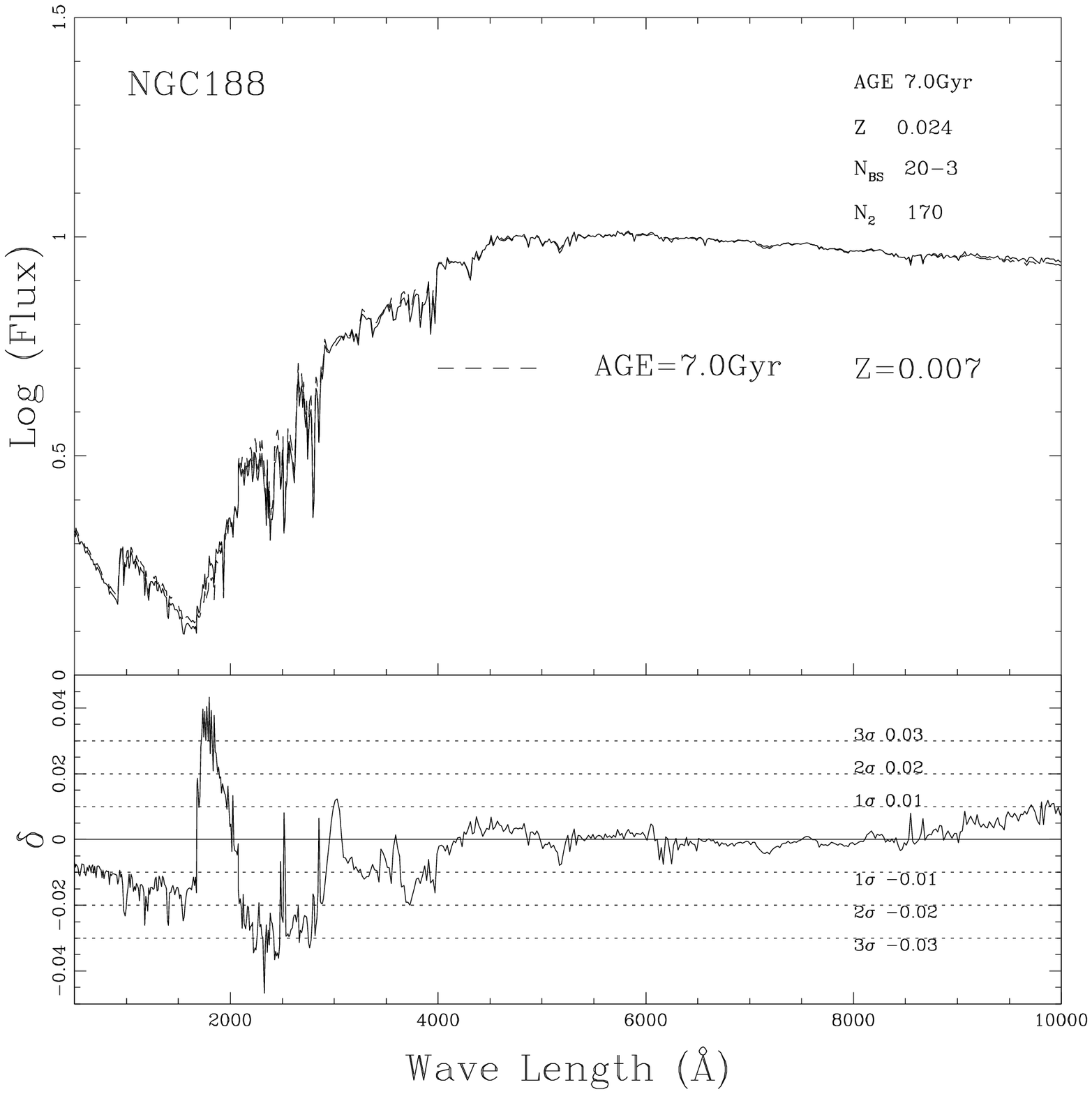}
\caption{Fitting the composite ISED of NGC 188 with 
conventional SSP model. The abscissa is the wavelength in angstrom. The 
ordinate is the Log(FLUX), which is the logarithmic value of the 
absolute flux of the ISED and is normalized at 5500 ${\AA}$. 
$\delta$ is the difference between the composite and conventional 
ISEDs, the direct subtraction of these two fluxes. Standard deviation
$\sigma$ is given in dotted lines in $\pm$3$\sigma$. 
The left panel is the fitting result of keeping the same metallicity 
but younger age, while the right panel is that just the opposite, 
keeping the same age but lower metallicity. 
The composite ISED is plotted in solid line. The conventional ISED
is given in dash line. The basic parameters of the cluster adopted 
in our work are listed in the top right corner in each panel. 
The parameters of the conventional ISED are given below the fitting 
line.\label{fig20}}
\end{figure}

\begin{figure}
\plottwo{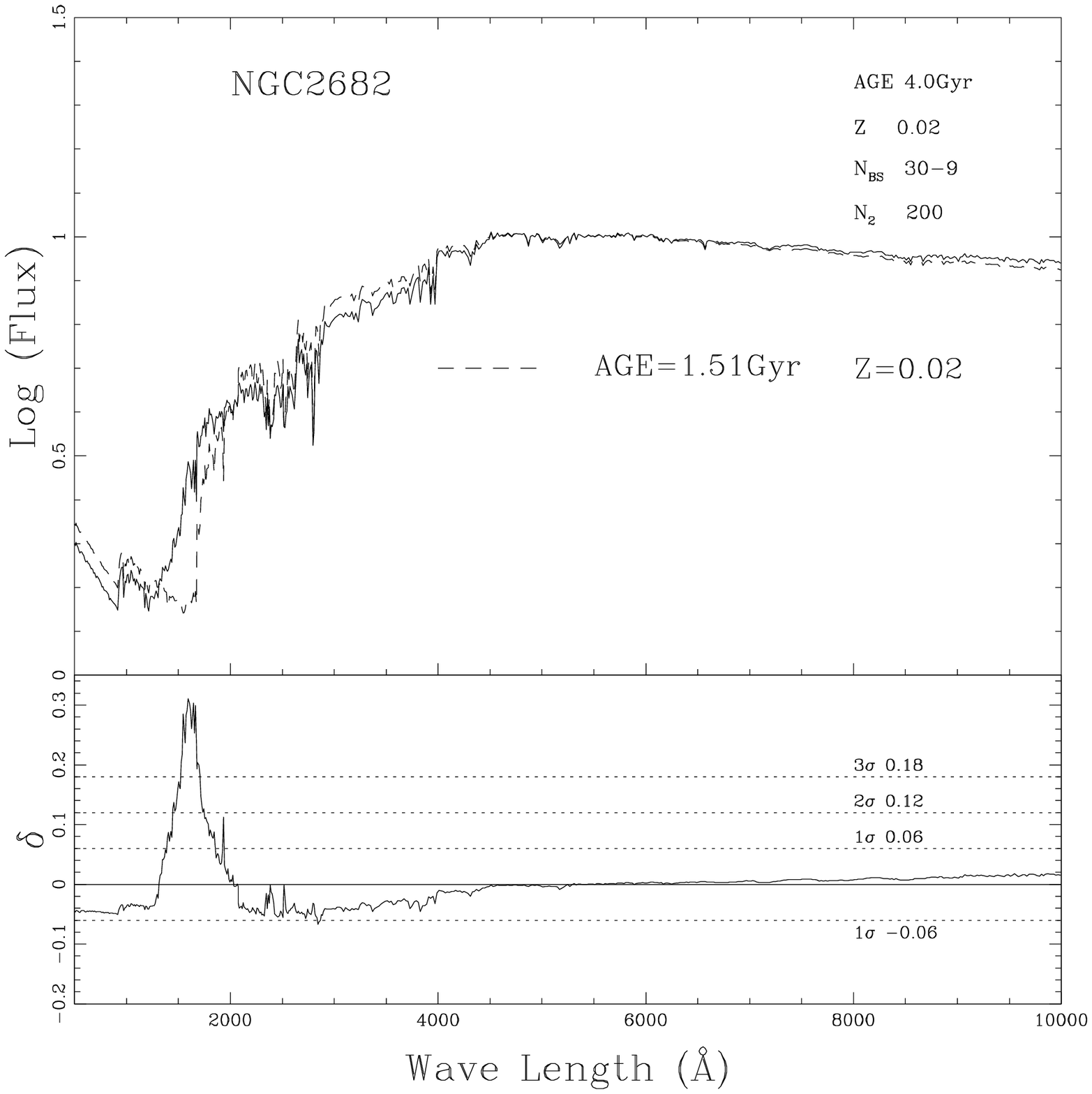}{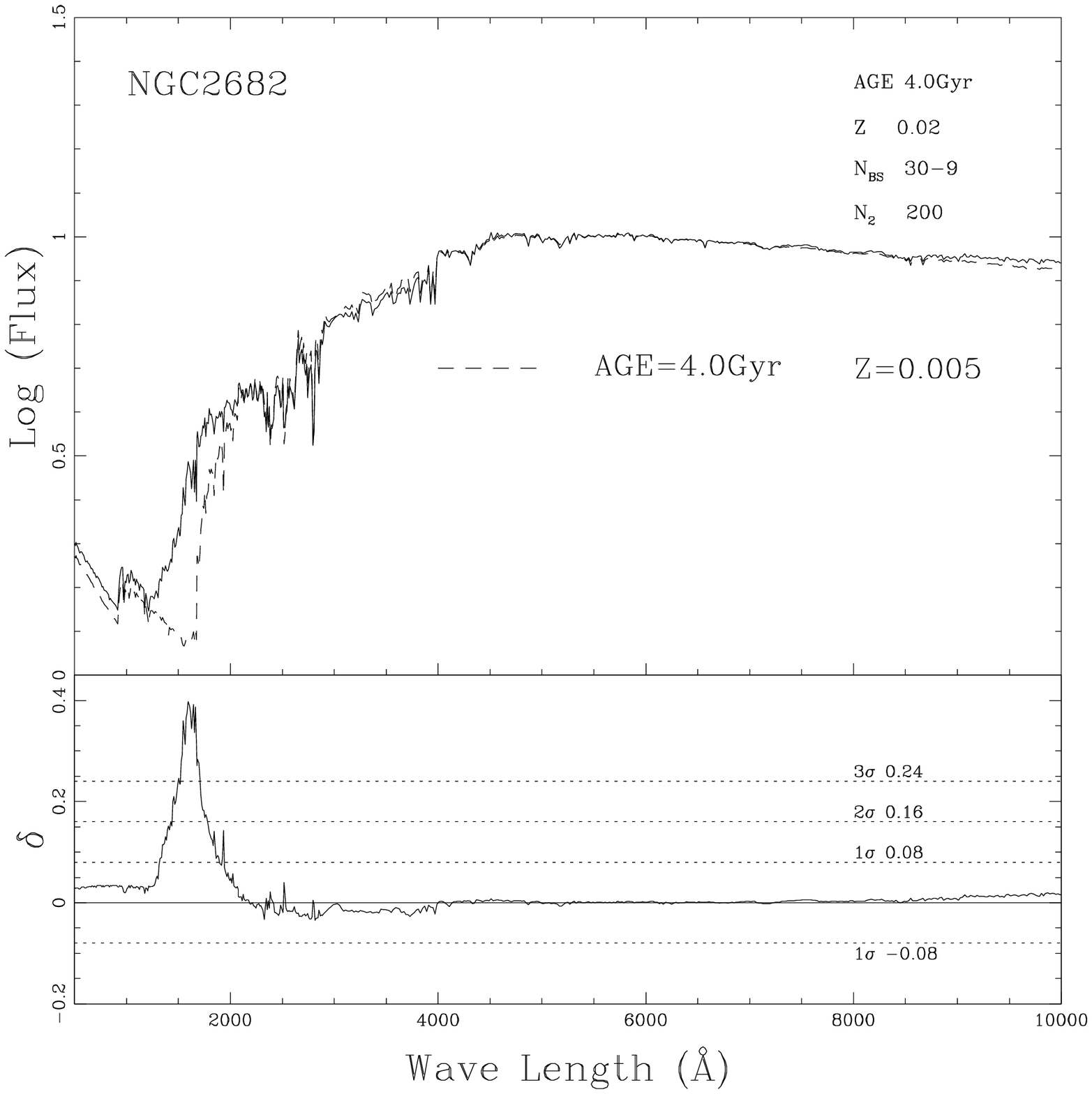}
\caption{Similar to Figure 20, but for NGC 2682.\label{fig21}}
\end{figure}

\end{document}